\documentclass[12pt,aps,pra,eqsecnum,subeqn,graphicx]{article}
\usepackage{amssymb,amsmath,color,times,epsfig}
\newcommand{\be}{\begin{equation}}
\newcommand{\ee}{\end{equation}}
\newcommand{\ba}{\begin{array}}
\newcommand{\ea}{\end{array}}
\newcommand{\beqa}{\begin{eqnarray}}
\newcommand{\eeqa}{\end{eqnarray}}
\newcommand{\beqas}{\begin{eqnarray*}}
\newcommand{\eeqas}{\end{eqnarray*}}
\newcommand{\beqal}{\begin{lefteqnarray}}
\newcommand{\eeqal}{\end{lefteqnarray}}

\begin{document}
\thispagestyle{empty} \baselineskip 0.7cm
\title{Determining neutralino parameters in Left-Right Supersymmetric Models}

\author{Nibaldo Alvarez-Moraga${}^{1,}$\thanks{{\it Email address:}
nibaldo.alvarez.m@exa.pucv.cl} \; and Artorix de la Cruz de
O\~na${}^{1,2,}$\thanks{{\it Email address:}
artorde@vax2.concordia.ca}
\\{\small \it  ${}^1$ Autonomous Center of theoretical Physics and Applied Mathematics,}
\\
{\small \it 3561 Hutchison $\#$ 3, Montr\'eal (Qu\'ebec)
H2X 2G9, Canada}\\
{\it ${}^2$ \small Department of Mathematic and Statistics,
Concordia University}, \\ {\it \small 1455 de Maisonneuve Blvb.
West, Montr\'eal (Qu\'ebec) H3G 1M8, Canada } }

 \maketitle

\begin{abstract}
We report exact analytical expressions relating  the fundamental
parameters describing the neutralino sector in the context of the
left-right supersymmetric model. The method used for such a effects
is the projector formalism deduced without take into account the
Jarlskog's projector formulae. Also, expressions for the neutralino
masses and the neutralino mixing matrix are determined . The results
are compared with numerical and analytical ones obtained in similar
scenarios in the context of the minimal supersymmetric standard
model.
\end{abstract}


 \vspace{5mm}

\setcounter{equation}{0}
\newpage
\section{INTRODUCTION}

In Ref. \cite{key1}, based on the Jarlskog's treatement of the
Cabibbo-Kabayashi-Maskawa matrix, the neutralino observables, in the
context of the minimal supersymmetric standard (MSSM),  were
described  in terms of projectors.  There, exact analytic
expressions for the neutralino masses were also obtained by
diagonalizing the associated real symmetric neutralino mass matrix.
Then, the same formalism was applied to treat a more general case
where the associated neutralino mass matrix was given by a complex
symmetric matrix Ref. \cite{key2}. In this last reference, several
CP conserving and violating possible scenarios were considered in
the study of the determining parameters of the theory.

The purpose of this work is  first to apply the projector formalism
\cite{key1,key2} to study  the existing connections among the
fundamental parameters describing the neutralino sector in the
contexte of the left-right supersymmetric (L-R SUSY) model
\cite{key3}-\cite{key4}. Next, to compare the results obtained to
the ones obtained in the contexte of the MSSM \cite{key2}.

In the L-R SUSY model which is  based on the gauge group $SU(2)_L
\times SU(2)_R \times U(1)_{B-L},$ the masses and mixing matrices of
the neutralinos and charginos are determined by $M_L$ , $M_R,$ the
left-right gaugino mass parameters associated with the gauge group
$SU (2)_L$ and $SU(2)_R$ respectively, $M_V,$ the gaugino mass
parameter associated with the gauge group $ U(1)_{B-L},$ $\mu,$ the
Higgsino mass parameter and the ratio $\tan\theta_k\equiv {k_{u}/
k_{d}},$ where $k_{u}$ and $k_d$ are the vacuum expectation values
of the Higgs fields which couple to $d$-type and $u$-type quarks
respectively \cite{key5}-\cite{key12}.

In Section \ref{sec-LRSUSY}, we give a brief description of the L-R
SUSY model and we write the Lagrangian density describing the
neutralino sector in terms  of the  two component fermion fields and
the neutralino mass matrix expressed in terms of the fundamental
parameters $M_L,$ $\mu,$ $\tan\theta_k,$ $ M_R $ and $M_V,$ where
$M_L$ and $\mu$ are considered, in general, as complex numbers. In
section \ref{sec-NEUTRALINO-MASS}, we compute the exact analytical
expressions for the neutralino masses and the corresponding
diagonalizing unitary matrix. Also, we plot these masses versus the
Higgsino parameter, in both the CP-conserving and CP-violating
cases, and we compare the corresponding CP-conserving results with
the  numerical ones obtained in \cite{key11}. In Section
\ref{sec-projectors}, the projector formalism \cite{key13} for this
model is revised. Based on the explicit construction of the
diagonalizing neutralino mass matrix, new formulas for the so-called
reduced projectors are constructed without appealing to the
Jarlskog's projector formulas \cite{key13,key14}. The fundamental
properties of these reduced projectors as well as the projectors and
the so-called pseudoprojectors \cite{key2} are proved. Also, the
equivalence of this reduced projectors with those obtained using the
Jarlskog's formulas is proved. In Section
\ref{sec-genral-ml-eigenphases}, using the new reduced projector
formulas, we express the complex parameter $M_L, $ in terms of the
so-called eigenphases \cite{key2} and the rest of the parameters.
 Moreover, taking advantage of the
mentioned equivalence we get a novel formula expressing the norm of
this complex parameter in terms of its phase and of the remaining
fundamental parameters. An alternative method to disentangle these
parameters are presented in Appendix \ref{sec-ml-appendix}.  In
Section \ref{sec-detparameters}, we compare the expected values of
the fundamental parameters in similar scenarios predicted by both
the L-R SUSY model and the MSSM. Finally, in Section
\ref{sec-conclusions}, we give our conclusions and prospects.

\section{A brief description of the Left-Right supersymmetric
model} \label{sec-LRSUSY} In the L-R SUSY model the full lagrangian
is given by \cite{key4} \be {\cal L} = {\cal L}_{\rm gauge} +  {\cal
L}_{\rm matter} + {\cal L}_{\rm Y} - {\cal V} + {\cal L}_{\rm soft},
\ee where ${\cal L}_{\rm gauge}$ contains the kinetic and
self-interactions terms for the bosons vector fields
$(W^\pm,W^0)_{L,R}$ and $ V^0,$ and the  Dirac Lagrangian of their
corresponding superpartners, i.e., the gaugino fields
$(\lambda^\pm,\lambda^0)_{L,R}$ and $\lambda^0_V;$ ${\cal L}_{\rm
matter}$ contains the kinetic terms for the fermionic and bosonic
matter fields, the Higgs fields and interaction of the gauge and
matter multiplets;  $\cal V$ is a scalar potential, ${\cal L}_{\rm
Y}$ (Yukawa Lagrangian) contains the self-interaction terms of the
matter multiplets as well as of the Higgs multiplets, e.g., it
contains  the self-interaction terms involving the fundamental
Higgsino mass parameters $\mu_1 \equiv \mu, \mu_2$ and $\mu_3:$ $
{\rm Tr} [ \mu_1 (\tau_1 {\tilde \phi}_u \tau_1)^T {\tilde \phi}_d
], $ $ {\rm Tr} [\mu_2 (\tau \cdot {\tilde \Delta}_L) (\tau \cdot
{\tilde \delta}_L) \phi_d ] $ and $ {\rm Tr} [\mu_3 (\tau \cdot
{\tilde \Delta}_R) (\tau \cdot {\tilde \delta}_R) \phi_d ],$ where
$\tau_j, \, j=1,2,3$ are the usual Pauli matrices, ${\tilde \phi}_d,
$ ${\tilde \Delta}_{L,R}$ and ${\tilde \delta}_{L,R}$ are the
superpartners of the bi-doublet field $\phi_d$  and  the four
triplet fields $\Delta_{L,R}$ and $\delta_{L,R},$ respectively,
which we will define soon afterward (in the following we will
consider $\mu_2=\mu_3=0$); and ${\cal L}_{\rm soft}$ is the
soft-breaking Lagrangian, involving the fundamental gaugino mass
parameters $M_L, M_R$ and $M_V,$ which gives Majorana mass to the
gauginos:  \beqa \nonumber {\cal L}_{\rm soft} &=& M_L
(\lambda^a_L \lambda^a_L + {\bar \lambda}^a_L {\bar \lambda}^a_L) \\
\nonumber &+& M_R (\lambda^a_R \lambda^a_R + {\bar \lambda}^a_R
{\bar \lambda}^a_R)
\\ &+& M_V (\lambda^0_V \lambda^0_V + {\bar \lambda}^0_V {\bar \lambda}^0_V).
\eeqa

The Higgs sector contains two bi-doublet fields,
\begin{eqnarray} \label{eq:bi-doublet}\phi_{u,\,d}=\left(\begin{array}{cc}
\phi^0_{1}&\phi^+_{1}\\
\phi^-_{2}&\phi^0_{2}
\end{array}\right)_{u,\,d}\,\equiv\,\left(\begin{array}{cc}
\frac{1}{2},\,\frac{1}{2},\,0\\
\end{array}\right),
\eeqa and four triplet fields, \beqa
\Delta_{L,\,R}=\left(\begin{array}{cc}
\frac{1}{\sqrt{2}}\,\Delta^{+}&\Delta^{++}\\
\Delta^{0}&-\frac{1}{\sqrt{2}}\,\Delta^{+}
\end{array}\right)_{L,\,R},\eeqa
and \beqa \delta_{L,\,R}=\left(\begin{array}{cc}
\frac{1}{\sqrt{2}}\,\delta^{+}&\delta^{++}\\
\delta^{0}&-\frac{1}{\sqrt{2}}\,\delta^{+}
\end{array}\right)_{L,\,R}.\eeqa
The Higgs $\Delta_{L,\,R}$ transform as $(1,0,2)$ and $(0,1,2)$
respectively. The triplet Higgs $\delta_{L,\,R}$ which transform as
$(1,0,-2)$ and $(0,1,-2)$ respectively, are introduced to cancel
anomalies in the fermionic sector that would otherwise occurs.

In order to generate mass for the gauge bosons we can choice the
vacuum expectation values of the Higgs fields in the form
\cite{key12} \beqa \label{vevdelta}\langle \Delta_{L} \rangle &=&
\langle \delta_{L,R}\rangle = 0, \qquad \langle \Delta_{R} \rangle =
\begin{pmatrix}
0 & 0 \cr \\ \upsilon_{R} & 0 \cr \end{pmatrix},
\\\label{vevkud} \langle \phi_{u} \rangle  &=& \begin{pmatrix}
k_{u}&0 \cr \\
0&0 \cr \end{pmatrix}, \qquad \langle \phi_{d}
\rangle=\begin{pmatrix}
0&0 \cr \\
0&k_{d} \end{pmatrix}. \label{eq:vacumm-expectation-values}\eeqa
Thus, in a first stage, the spontaneous   breaking of
$SU(2)_{R}\times U(1)_{B-L}$ to $U(1)_Y,$ according to the vacuum
expectation value $\langle \Delta_R \rangle \ne 0,$ given in Eq.
\eqref{vevdelta}, generates masses for $W^{\pm}_R, W^0_R$ and $V^0.$
The two neutral states $W^0_R$ and $V^0 $ mix  yielding the physical
field $Z_R$ and the massless field  $B.$ The vacuum expectation
value $v_{R}$ of the triplet Higgs $\Delta_{R}$  has  being chosen
very big  to provide  large masses to gauge bosons $W^{\pm}_{R}$ and
$Z_R$. Next, through the spontaneous breaking of $SU(2)_{L}\times
U(1)_{Y}$ into $U(1)_{\rm em},$ according the chosen  vacuum
expectation values $\phi_{u,d}$ given in  Eq. \eqref{vevdelta},  the
left weak bosons $W^\pm_{L}$ and $W^0_L$ as well as  $B_\mu$ acquire
mass. Once again, the neutral fields mix forming the massless photon
$A_\mu$ and the physical gauge field $Z_L.$ The masses of the
right-handed gauge bosons are given by \be M_{W_R}= {1 \over
\sqrt{2}} g_R (k_u^2 + k_d^2 + v_R^2)^{1/2},\ee \be M_{Z_{R}}= {1
\over \sqrt{2}} v_R (g_R^2 + 4 g_V^2)^{1/2}, \ee whereas the  mass
of the left-handed ones are given by  \be M_{W_L}= {1\over \sqrt{2}}
g_L (k_u^2 + k_d^2)^{1/2}, \ee  \be M_{Z_{L}}= {1 \over \sqrt{2} } [
(k_u^2 + k_d^2) (g_L^2 + 4 {g^\prime}^2)]e^{1/2},\ee where
$g_{L},\,g_{R}, g_{V}$ and  $g'=
g_{R}\,g_{V}/(g^2_{R}+4g^2_{V})^{1/2}$ are the coupling constants of
the gauge groups $SU(2)_{L},$ $SU(2)_{R},$ $U(1)_{B-L}$ and $U
(1)_Y,$ respectively.

To find the neutralino masses  we must consider the interaction
terms between the gauge bosons, the Higgs, and their superpartners.
The neutralino particles are produced in two stages of  symmetry
breaking. The first stage involving  the vacuum expectation value
$v_R$ of $\Delta_R $ generates masses for three heavy neutralinos
$\tilde{\chi}^0_{k}, \, k=5,6,7. $  The second stage involving the
vacuum expectation values $k_u$ and $k_d$ of the Higgs $\phi_u$ and
$\phi_d$  generate mass for the light neutralinos
$\tilde{\chi}^0_{k}, \, k=1,\ldots,4.$  The Lagrangian for light
neutralinos is given by \cite{key12}  \beqa \label{eq:BMz}
\mathcal{L}_{LN} &=& \nonumber
-\frac{i}{\sqrt{2}}\,g_{L}\,k_{u}\,\tilde{\phi}^0_{1u}\,\lambda^{0}_{L}+i\,\sqrt{2}\,\frac{g_{R}\,g_{V}}{g_{1}}\,k_{u}\,\widetilde{\phi}^{0}_{1u}\,\lambda^{0}_{B}
\\ \nonumber &-& i\,\sqrt{2}\,\frac{g_{R}\,g_{V}}{g_{1}}\,k_{d}\,\widetilde{\phi}^{0}_{2d}\,\lambda^{0}_{B}+\frac{i}{\sqrt{2}}\,g_{L}\,k_{d}\,\widetilde{\phi}^{0}_{2d}
\\ \nonumber &+& M_{L}\,\lambda^{0}_{L}\,\lambda^{0}_{L} +
[\frac{(4\,M_{R}\,g^2_{V}+\,M_{V}\,g^2_{R})}{g_{1}}]\,\lambda^{0}_{B}\,\lambda^{0}_{B}
\\ &+&  2\mu\,
\widetilde{\phi}^{0}_{1u}\,\widetilde{\phi}^{0}_{2d}+\mathrm{h.c.},
\eeqa where $\lambda^0_B = (g_R \lambda^0_V + 2 g_V \lambda^0_R) /
g_1$ with $g_1= (g_R^2 + 4 g_V^2)^{1/2};$ $\lambda^0_{L,R}$ and
$\lambda^0_V$ are the neutral gaugino fields; and
$\widetilde{\phi}^{0}_{1u}$ and $\widetilde{\phi}^{0}_{2d}$ are the
neutral Higgsino fields, i.e., the superpartner of the neutral Higgs
fields ${\phi}^{0}_{1u}$ and ${\phi}^{0}_{2d},$ respectively,
defined in Eq. \eqref{eq:bi-doublet}.

The above Lagrangian in matrix form  can be written as follows \beqa
\label{eq:REM} \mathcal{L}_{LN}=-\frac{1}{2}\,(\xi^0)^T\, N
\,\xi^0+\mathrm{h.c.}, \eeqa
where $N$ is in general a complex symmetric matrix given by
\be \label{eq:Imatrix}  N=   \left(\begin{array}{cccc}
M_{L} & 0 & -\frac{1}{\sqrt{2}}\,g_{L}\,k_{u} & \frac{1}{\sqrt{2}}\,g_{L}\,k_{d}\\
0 & \frac{4\,M_{R}\,g^2_{V}+M_{V}\,g^2_{R}}{g^2_{1}} & \frac{\sqrt{2}\,g_{R}\,g_{V}\,k_{u}}{g_{1}} & -\frac{\sqrt{2}\,g_{R}\,g_{V}\,k_{d}}{g_{1}}\\
-\frac{1}{\sqrt{2}}\,g_{L}\,k_{u} & \frac{\sqrt{2}\,g_{R}\,g_{V}\,k_{u}}{g_{1}} & 0 & -2\mu\\
\frac{1}{\sqrt{2}}\,g_{L}\,k_{d} &
-\frac{\sqrt{2}\,g_{R}\,g_{V}\,k_{d}}{g_{1}}  & -2\mu & 0
\end{array}\right),
\ee   and the two component fermion field is \be\label{eq:U2}
(\xi^{0})^{T}=(-i\,\lambda^0_{L}\,\,\,-i\,\lambda^0_{B}\,\,\,\,\,\,\tilde{\phi}^0_{1u}\,\,\,\,\,\,\tilde{\phi}^0_{2d}).
\ee
%

\section{The neutralino masses and the diagonalizing matrix in the left-right supersymmetric model}
\label{sec-NEUTRALINO-MASS}

The two-component light neutralino mass eigenstates ${\chi}^{0}_j$
are related to the two component fermion fields given in Eq.
\eqref{eq:U2} as \be \label{eq:BMz} { \xi}^0_{k}= \sum_{l=1}^{4}
V_{k\,l} \, {\chi}^0_{l}, \qquad k=1,\ldots,4, \ee where $V$ is a
unitary matrix satisfying \beqa
 N_{D} &=& V^T \,N \,V, \nonumber \\
&\equiv& \sum_{j=1}^{4}\, m_{{\tilde \chi}^{0}_{j}}\,E_{j},
\label{eq:VMV} \eeqa and
\beqa
\label{eq:MD2} N^{2}_{D} &=& V^{-1}\, N^{\dag} \,N\,V,  \\
&\equiv& \sum_{j=1}^{4}\,m_{{\tilde \chi}^{0}_{j}}^{2}\,E_{j}, \eeqa
where $(E_{j})_{4\times4}$ are the basic matrices defined by
$(E_{j})_{ik}=\delta_{ji}\,\delta_{jk}$ and ${\tilde \chi}^0_j$
stand for the four component Majorana neutralinos:   \be {\tilde
\chi}^0_j =
\begin{pmatrix} \chi^0_j \cr \\ {\bar \chi}^0_j  \cr \end{pmatrix}, \qquad j=1,\ldots,4. \ee
Here, we suppose that the real eigenvalues of $N_{D}$ are ordered in
the following way \be \label{eq:orden-masas}  m_{{\tilde
\chi}^{0}_{1}} \le   m_{{\tilde \chi}^{0}_{2}}  \le  m_{{\tilde
\chi}^{0}_{3}} \le  m_{{\tilde \chi}^{0}_{4}} .\ee

\subsection{Exact analytical expressions for the neutralino masses}
As we have seen in the above section, in the left-right
supersymmetric model, the masses, the mixing parameters and the
CP-violating properties of the neutralino are determined by the
fundamental complex $M_L= |M_L| e^{i \Phi_L }$ and $\mu= |\mu| e^{i
\Phi_\mu }$ and real $tan\, \theta_k = k_u / k_d, $ $M_R $ and $M_V$
parameters. To know the neutralino masses predicted  by the present
model, we can solve the characteristic equation associated to the
Hermitian matrix $ H\equiv N^{\dag} \,N$. More precisely, the square
root of the positive roots of this characteristic equation
corresponds to the physical neutralino masses. The neutralino masses
predicted by the present model are known only for the  CP-conserving
case under  the limit of large $M_{L,R}$ or large $|\mu|$
\cite{key12}, more precisely on the assumptions that $|M_{L,R} \pm
\mu | \gg M_{Z_L},$ $M_R > M_V,$ and $4 g_V^2 M_R + g_R^2 M_V /
g_1^2 \simeq 4 g_V^2 M_R /g_1^2.$ Indeed, a numerical analysis has
been implemented to solve the mentioned characteristic equation
\cite{key12}, assuming determined values for the gauge boson masses,
couplings constants and taking $\mu,$ the higgsino mass parameter,
as a free quantity. Here, we put into practice a method
\cite{key1,key2} giving exact analytic expressions for the
neutralino masses.

Starting from Eq. \eqref{eq:MD2}, we get  \be \label{eq:EVP}
(N^{\dag}\,N) \, V  - V \, N^{2}_{D} =0. \ee A more explicit form of
this matrix equation is \beqa \label{eq:FE-EVP}
(H_{11}- m_{{\tilde \chi}^{0}_{j}}^2)V_{1j}+H_{12}V_{2j}+H_{13}V_{3j}+H_{14}V_{4j}\nonumber &=& 0,\\
H_{21}V_{1j}+ (H_{22}-m_{{\tilde \chi}^{0}_{j}}^2)V_{2j}+H_{23}V_{3j}+H_{24}V_{4j}\nonumber &=& 0,\\
H_{31}V_{1j}+H_{32}V_{2j}+(H_{33}-m_{{\tilde \chi}^{0}_{j}}^2)V_{3j}+H_{34}V_{4j}\nonumber &=& 0,\\
H_{41}V_{1j}+H_{42}V_{2j}+H_{43}V_{3j}+(H_{44}-m_{{\tilde
\chi}^{0}_{j}}^2)V_{4j} &=& \nonumber 0, \\
\eeqa $j=1,\ldots, 4,$ where $H_{ij}= \sum_{k=1}^4 N^{\ast}_{ki}
N_{kj}:$

\beqa \nonumber H_{11}&=& M^2 + |M_L|^2, \\ \nonumber H_{22} &=& 4
\kappa^2 M^2 + M_{RV}^2,
\\ \nonumber H_{33}&=& 4 {|\mu|}^2 +  (1
+ 4 \kappa^2) M^2 \sin^2 \theta_k,
\\ \nonumber H_{44}&=& 4 {|\mu|}^2 + (1
+ 4 \kappa^2) M^2 \cos^2 \theta_k,
\\\nonumber H_{12}&=& H_{21}^\ast =  - 2 \kappa  M^2
\\
\nonumber H_{13}&=& H_{31}^\ast =  -  M \, \bigl( 2 |\mu| e^{i
\Phi_\mu} \cos\theta_k  + |M_L| e^{-i \Phi_L} \sin\theta_k \bigr),
\\
\nonumber H_{14}&=& H_{41}^\ast =   M \, \bigl( 2 |\mu| e^{i
\Phi_\mu} \sin\theta_k  + |M_L| e^{-i \Phi_L} \cos\theta_k\bigr),
\\
\nonumber H_{23}&=& H_{32}^\ast =2 \kappa  M \, \bigl( 2 |\mu| e^{i
\Phi_\mu} \cos\theta_k + M_{RV} \sin\theta_k \bigr),
\\
\nonumber H_{24}&=& H_{42}^\ast =- 2 \kappa M \, \bigl( 2 |\mu| e^{i
\Phi_\mu} \sin\theta_k  + M_{RV} \cos\theta_k\bigr),
\\
\nonumber H_{34}&=& H_{43}^\ast =  - { 1 \over 2}  (1 + 4 \kappa^2)
M^2 \sin ( 2 \theta_k ), \eeqa where $M=  g_L M_{Z_L} / \sqrt{g_L^2
+ 4 {g^\prime}^2 },$  $\kappa= g_R g_V /  g_1 g_L,$ and
 $M_{RV}= (4 g_V^2 M_R + g_R^2 M_V) / g_1^2.$

For fixed  $j,$  Eq. \eqref{eq:FE-EVP} represents a  system of
homogeneous linear equations depending on  only one of the
neutralino masses. Thus, the neutralino masses can be determined by
solving the characteristic equation associated to this system, that
is \be
 X^{4} - a\,X^{3}
+  b\,X^{2} - c\, X + d=0\,, \label{eq:CEQ} \ee where \beqas
\label{eq:a-term} \nonumber a &=& |M_L|^2 +   8 |\mu|^2  +  M_{RV}^2
+ 2  (1 + 4 \kappa^2) M^2,  \eeqas  \beqas \nonumber b &=& \bigl( 4
|\mu|^2 + (1 + 4 \kappa^2) M^2 \bigr)^2
\\ \nonumber  &+&  M_{RV}^2 \bigl( |M_L|^2 + 8 |\mu|^2 + 2 M^2
\bigr) \\ \nonumber & +&
 8 |M_{L}|^2 \left(|\mu|^2 +  \kappa^2 M^2 \right) \\
 \nonumber &
 -& 16 \kappa^2 M^2 |\mu| M_{RV} \sin(2 \theta_k ) \cos \Phi_{\mu}
\\\nonumber &-& 4 M^2 |\mu| |M_{L}| \sin(2 \theta_k ) \cos (\Phi_{\mu} +
\Phi_L), \eeqas \beqas \nonumber c &=&  16 |\mu|^4 |M_L|^2 +  4 (1 +
4 \kappa^2)^2 M^4 |\mu|^2 \sin^2 ( 2 \theta_k)
\\ &+&  16  \kappa^2 M^2 |M_L|^2
\left(2 |\mu|^2  + \kappa^2 M^2 \right) \\
&+& M_{RV}^2 \bigl( M^4 + 8 |\mu|^2 (M^2 + |M_L|^2)  + 16 |\mu|^4
\bigr) \\ &-& 4 M^2 |\mu| |M_L| \bigl(
 4 |\mu|^2 + M_{RV}^2 \bigr) \cos(\Phi_L + \Phi_\mu) \sin(2
\theta_k)\\ &+& 8 \kappa^2 M^2 M_{RV}
 \bigl[ M^2 |M_L| \cos\Phi_L \\&-& 2 |\mu| (4 |\mu|^2 +
|M_L|^2 ) \cos\Phi_\mu \sin(2 \theta_k ) \bigr], \eeqas  and \beqas
\nonumber
 d &=& 64 \kappa^4
M^4 |\mu|^2 |M_L|^2 \sin^2 (2\theta_k) \\ &+& 32 \kappa^2 M^2
|\mu|^2 |M_L| M_{RV} \sin(2 \theta_k)
\\ &\times& \bigl( M^2 \cos\Phi_L \sin(2 \theta_k) - 2 |\mu| |M_L|
\cos\Phi_\mu \bigr) \\ &+& 4 M^2 |\mu|^2 M_{RV}^2 \sin(2 \theta_k) \\
&\times& \bigl(M^2 \sin(2 \theta_k) - 4 |\mu| |M_L| \cos(\Phi_L +
\Phi_\mu ) \bigr) \\ &+& 16   |\mu|^4 |M_L|^2 M_{RV}^2. \eeqas
Solving Eq. \eqref{eq:CEQ}, we get the exact analytic formulas for
the neutralino  masses \beqa m_{{\tilde \chi}^{0}_{1}}^2 ,
m_{{\tilde \chi}^{0}_{2}}^2 &=& \frac{a}{4}-\frac{\alpha}{2}\mp\frac{1}{2}\,\sqrt{\beta - \varpi-\frac{\lambda}{4\alpha}},\\
m_{{\tilde \chi}^{0}_{3}}^2,m_{{\tilde \chi}^{0}_{4}}^2 &=&
\frac{a}{4}+\frac{\alpha}{2}\mp\frac{1}{2}\,\sqrt{\beta
-\varpi+\frac{\lambda}{4\alpha}},\label{eq:EIGV} \eeqa
where \beqa \alpha
&=& \nonumber \sqrt{{\beta \over 2} + \varpi},\\
\nonumber \varpi &=&\frac{\epsilon}{3 \ 2^{\frac{1}{3}}} +
\frac{(2^{\frac{1}{3}}\,\gamma)}{3\,\epsilon},\\ \nonumber \epsilon
&=& (\delta+ \sqrt{\delta^2 - 4 \gamma^3})^{\frac{1}{3}},
\\ \nonumber \beta &=& \frac{a^{2}}{2}-\frac{4b}{3},\\\nonumber
 \lambda &=& \nonumber a^{3}- 4\,a\,b + 8\,c\\ \nonumber
\gamma &=& \nonumber b^{2}- 3\,a\,c+12\,d, \\ \nonumber \delta &=&
\nonumber 2\,b^{3}- 9\,a\,b\,c+27\,c^{2}+ 27\,a^{2}\,d-72\,b\,d.
\eeqa

\subsection{Neutralino masses, numerical results}
\begin{table}
\begin{center}
\begin{tabular}{c c c c c }\cline{1-5}\\
Scenario & $M_R \; $ & $ M_L \; $ & $ k_u $ & $ \tan\theta_k  $ \\
\\\cline{1-5} \\
$Sc_1 $ & 300 & 50  &  92.75 &  \parbox[c]{0.5cm}{ 1.6    \\
4.0} \\ \\  \hline \\
$Sc_2 $ & 1000 & 250  &  92.75 &  \parbox[c]{0.5cm}{ 1.6  \\
4.0}
\\ \\  \hline \\
\end{tabular}
\end{center}
\caption{Input parameters for scenarios $Sc_1$ and $Sc_2.$  All mass
quantities are given in GeV.} \label{tab:tablauno}
\end{table}

Let us consider the CP-conserving scenarios $Sc_1$ and $Sc_2$
described in Tab. \ref{tab:tablauno}. This scenarios are similar to
the ones studied in Ref. \cite{key12} where they have been used to
compare the predicted results for the neutralino masses in the
left-right SUSY model and the MSSM.  Thus, for both scenarios, we
consider the coupling constants values $g_R\approx g_L \approx g_V =
0.65,$ the gaugino parameters $M_L > >  M_V \approx 0.0 $GeV and the
mixing phases $\Phi_L=\Phi_\mu=0.$ Figures \ref{fig:masses1} and
\ref{fig:masses2}, show the behavior of the physical neutralino
masses $m_{{\tilde \chi}^{0}_{i}}, \, i=1,\ldots,3,$ versus $\mu,$
computed from Eq. \eqref{eq:EIGV}, for the inputs of scenarios
$Sc_1$ and $Sc_2$ with $\tan\theta_k=1.6,$ respectively. Notice that
the values of the  neutralino mass $m_{{\tilde \chi}^{0}_{4}}$ are
so big that they cannot be seen. Both Figures reproduce accurately
the results of Ref. \cite{key12}. We observe the correct size
ordering of the neutralino masses, such as required by Eq.
\eqref{eq:orden-masas}. Also, in both  scenarios,  we find that for
values of $ |\mu| \sim 200 $GeV, the neutralino masses $m_{{\tilde
\chi}^{0}_{1}}$ are approximately   $M_L$  and for large values of
$|\mu|,$ the masses of the neutralinos $m_{{\tilde \chi}^{0}_{i}},
i= 3,4, $ are heavier than $ M_R.$ The same analysis is true in the
case of scenarios $Sc_1$ and $Sc_2,$ where $\tan\theta_k=4.0,$ as we
can observe in Figs. \ref{fig:masses3} and \ref{fig:masses4}.
However, comparing Figs. \ref{fig:masses1} and \ref{fig:masses3},
corresponding to scenarios $Sc_1$ with different values of
$\tan\theta_k$, i.e.,
 $\tan \theta_k=1.6$ and   $\tan \theta_k=4.0,$ respectively, we find
 that for small values of $|\mu|,$ the variation
 of the  neutralino masses with respect to $\mu$ in Fig. \ref{fig:masses3}
 are smoother than in  Fig. \ref{fig:masses1}. This is an important fact to consider when
 we will study the inverse problem, that is,  the determination
 of  the fundamental parameters  based on the knowledge of the physical neutralino
 masses.

\begin{table}
\begin{center}
\begin{tabular}{c c c c c c }\cline{1-6}\\
Scenario & $|\mu| $\; &  $M_R \; $ & $ M_L \; $ & $ k_u $ & $ \tan\theta_k  $ \\
\\\cline{1-6} \\
$Snc_1 $ & \parbox[c]{0.5cm}{20 \\ 248} & 300 & 50  &  92.75 &   4.0  \\ \\  \hline \\
\end{tabular}
\end{center}
\caption{Input parameters for scenario $Snc_1.$   All mass
quantities are given in GeV.} \label{tab:tablatres}
\end{table}

\begin{figure} \centering
\begin{picture}(31.5,21)
\put(1,2){\includegraphics[width=70mm]{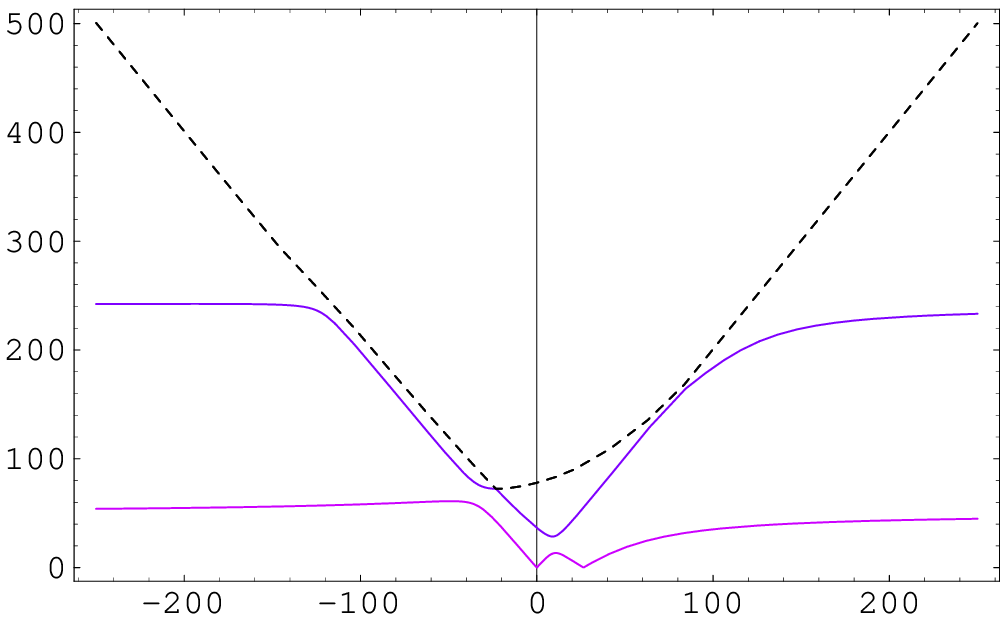}}
\put(15.75,0.7){\scriptsize{$\mu \, {\rm (GeV)}$}}
\put(1.05,19.25){\scriptsize{$m_{{\tilde \chi}^{0}_{i}} \, {\rm
(GeV)}$}}
\end{picture}
\caption{Neutralino masses $m_{{\tilde \chi}^{0}_{i}}, \,
i=1,\ldots,3,$ as functions of $\mu$ for scenario $Sc_1,$ assuming $
\tan \theta_k = 1.6.$} \label{fig:masses1}
\end{figure}

\begin{figure} \centering
\begin{picture}(31.5,21)
\put(1,2){\includegraphics[width=70mm]{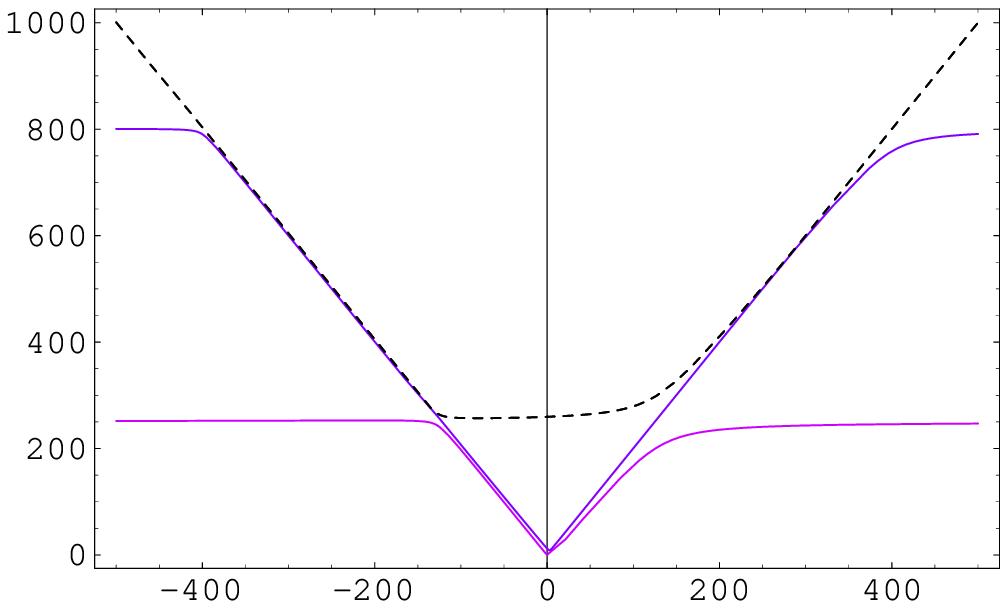}}
\put(15.75,0.7){\scriptsize{$\mu \, {\rm (GeV)}$}}
\put(1.05,19.25){\scriptsize{$m_{{\tilde \chi}^{0}_{i}}\, {\rm
(GeV)} $}}
\end{picture}
\caption{Neutralino masses $m_{{\tilde \chi}^{0}_{i}}, \,
i=1,\ldots,3,$ as functions of $\mu$ for scenario $Sc_2,$ assuming $
\tan \theta_k = 1.6.$ }\label{fig:masses2}
\end{figure}

Let us now to study the behavior of the neutralino masses
$m_{{\tilde \chi}^{0}_{i}}, i=1,2,$  respect to the variation of
$\Phi_\mu$ and $\Phi_L.$ Let us consider two possible CP-violating
scenarios $Snc_1$ described in Tab. \ref{tab:tablatres},
characterized by  two different values of  the Higgsino mass
parameter,  $|\mu|=20$GeV and $|\mu|=248$GeV. Figures
\ref{fig:mass1_u=20_FU_FIL} and \ref{fig:mass2_u=20_FU_FIL} show the
behavior of the  neutralino masses $m_{{\tilde \chi}^{0}_{1}} $  and
$m_{{\tilde \chi}^{0}_{2}}, $ respectively, as a function of
$\Phi_\mu$ and $ \Phi_L$  for input parameters of scenario $Snc_1$
with $|\mu|=20$GeV. Comparing these figures we observe that the
variation of the values of $m_{{\tilde \chi}^{0}_{1}} $ is bigger
than the one of $m_{{\tilde \chi}^{0}_{2}}. $ Superposing these
figures, the corresponding surfaces do not overlap , that is, the
size ordering (see Eq. \eqref{eq:orden-masas}) of the masses
$m_{{\tilde \chi}^{0}_{1}} $ and $m_{{\tilde \chi}^{0}_{2}},$ is
conserved even if in the CP-violating case.  Figures
\ref{fig:mass1_u=248_FU_FIL} and \ref{fig:mass2_u=248_FU_FIL} show
the behavior of the neutralino masses $m_{{\tilde \chi}^{0}_{1}} $
and $m_{{\tilde \chi}^{0}_{2}}, $ respectively,  as a function of
$\Phi_\mu$ and $ \Phi_L$  for input parameters of scenario  $Snc_1$
with $|\mu|=248$GeV. The same considerations as in the previous
analysis done for Figs. \ref{fig:mass1_u=20_FU_FIL} and
\ref{fig:mass2_u=20_FU_FIL} are valid in this case. However, we
observe that the energy gap between the surfaces in Figs.
\ref{fig:mass1_u=248_FU_FIL} and \ref{fig:mass2_u=248_FU_FIL} is
greater than in the case of surfaces of Figs.
\ref{fig:mass1_u=20_FU_FIL} and \ref{fig:mass2_u=20_FU_FIL}.

\begin{figure} \centering
\begin{picture}(31.5,21)
\put(1,2){\includegraphics[width=70mm]{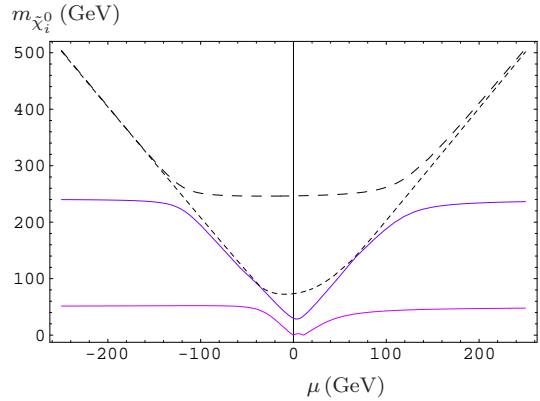}}
\put(15.75,0.7){\scriptsize{$\mu \, {\rm (GeV)}$}}
\put(1.05,19.25){\scriptsize{$m_{{\tilde \chi}^{0}_{i}}\, {\rm
(GeV)} $}}
\end{picture}
\caption{Neutralino masses $m_{{\tilde \chi}^{0}_{i}}, \,
i=1,\ldots,4,$ as functions of $\mu$ for scenario $Sc_1,$ assuming $
\tan \theta_k = 4.0.$} \label{fig:masses3}
\end{figure}

\begin{figure} \centering
\begin{picture}(31.5,21)
\put(1,2){\includegraphics[width=70mm]{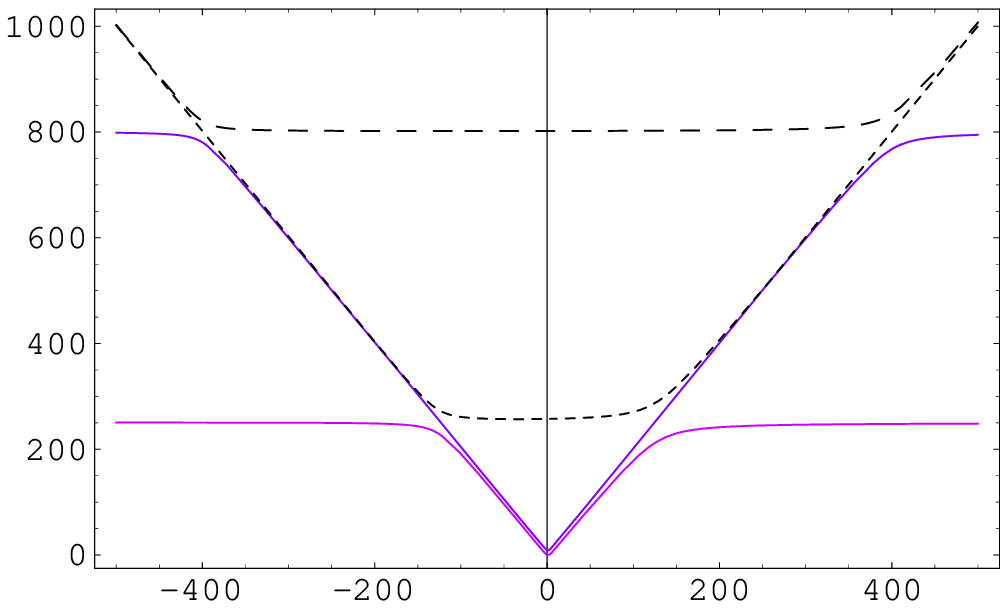}}
\put(15.75,0.7){\scriptsize{$\mu \, {\rm (GeV)}$}}
\put(1.05,19.25){\scriptsize{$m_{{\tilde \chi}^{0}_{i}} \, {\rm
(GeV)}$}}
\end{picture}
\caption{Neutralino masses $m_{{\tilde \chi}^{0}_{i}}, \,
i=1,\ldots,4,$ as functions of $\mu$ for scenario $Sc_2,$ assuming $
\tan \theta_k = 4.0.$} \label{fig:masses4}
\end{figure}


\begin{figure} \centering
\begin{picture}(36.5,25)
\put(7,2){\includegraphics[width=70mm]{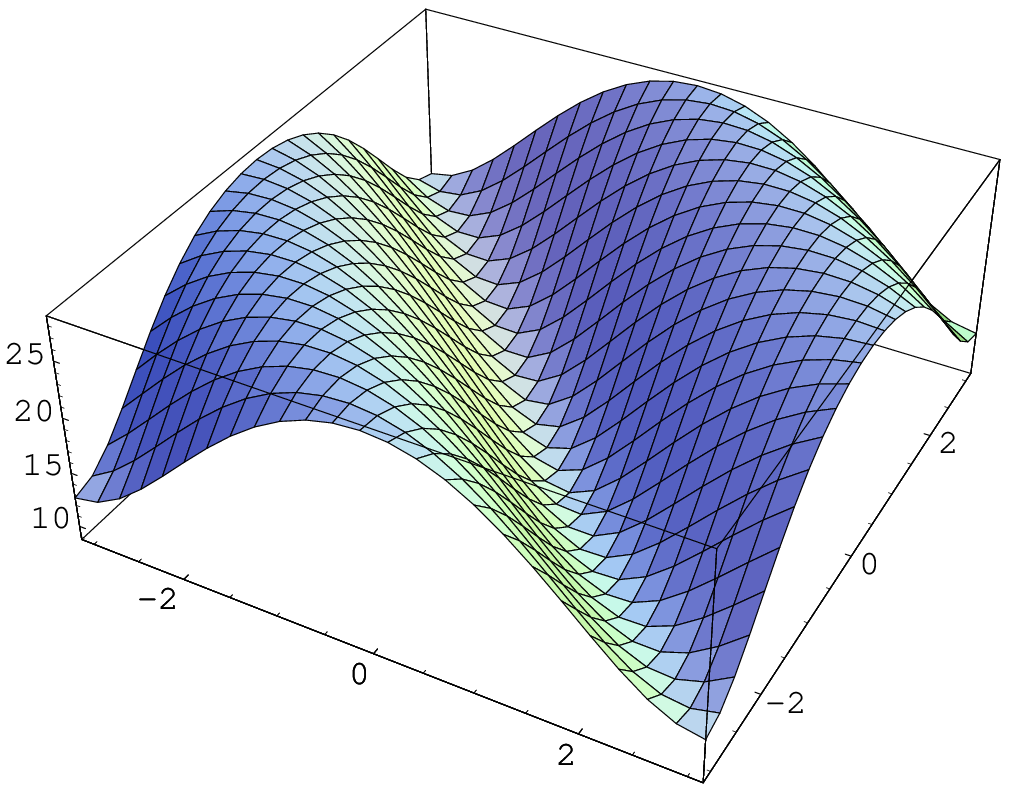}}
\put(13,3){\scriptsize{$\Phi_\mu \, {\rm (rad)}$}}
\put(30,6){\scriptsize{$\Phi_L \, {\rm (rad)}$}}
\put(1,10){\scriptsize{$m_{{\tilde \chi}^{0}_{1}} \, {\rm (GeV)}$}}
\end{picture}
\caption{Neutralino masse $m_{{\tilde \chi}^{0}_{1}}, $ as a
function of  $\Phi_\mu$ and $ \Phi_L$  for input parameters
according to  scenario $Snc_1$ with  $|\mu|=20$GeV.}
\label{fig:mass1_u=20_FU_FIL}
\end{figure}

\begin{figure} \centering
\begin{picture}(36.5,25)
\put(7,2){\includegraphics[width=70mm]{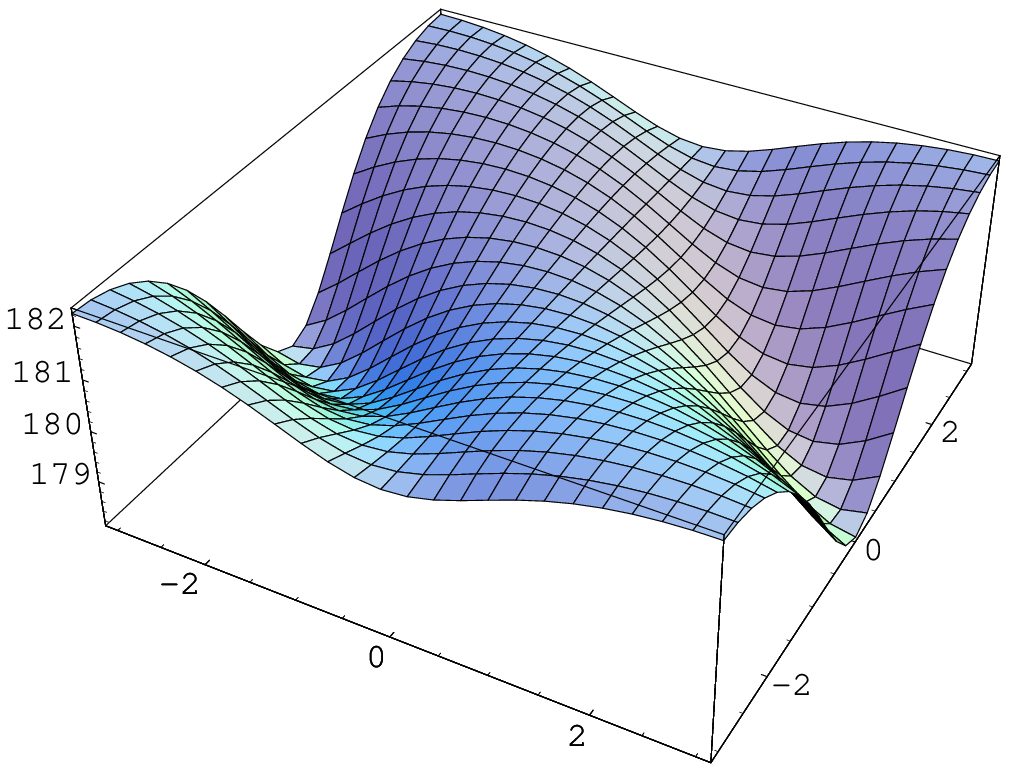}}
\put(13,3){\scriptsize{$\Phi_\mu \, {\rm (rad)}$}}
\put(30,6){\scriptsize{$\Phi_L \, {\rm (rad)}$}}
\put(1,10){\scriptsize{$m_{{\tilde \chi}^{0}_{1}} \, {\rm (GeV)}$}}
\end{picture}
\caption{Neutralino masse $m_{{\tilde \chi}^{0}_{2}}, $ as a
function of  $\Phi_\mu$ and $ \Phi_L$  for input parameters
according to scenario $Snc_1$ with  $|\mu|=20$GeV.}
\label{fig:mass2_u=20_FU_FIL}
\end{figure}

\begin{figure} \centering
\begin{picture}(36.5,25)
\put(7,2){\includegraphics[width=70mm]{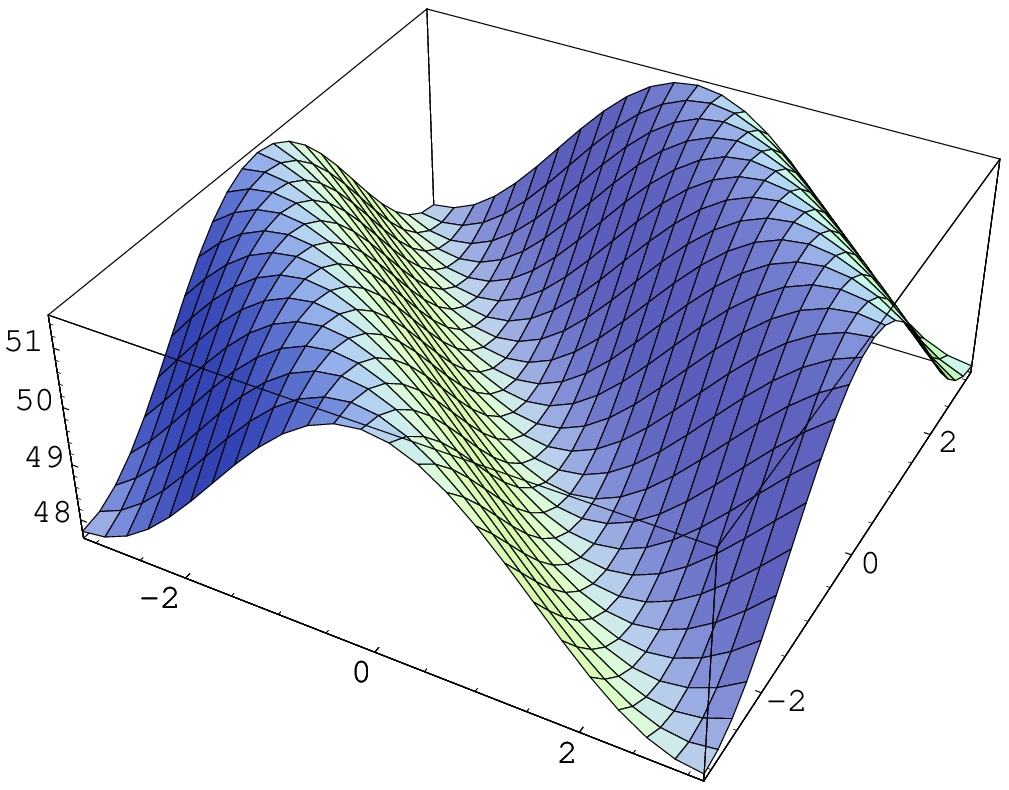}}
\put(13,3){\scriptsize{$\Phi_\mu \, {\rm (rad)}$}}
\put(30,6){\scriptsize{$\Phi_L \, {\rm (rad)}$}}
\put(1,10){\scriptsize{$m_{{\tilde \chi}^{0}_{1}} \, {\rm (GeV)}$}}
\end{picture}
\caption{Neutralino masse $m_{{\tilde \chi}^{0}_{1}}, $ as a
function of  $\Phi_\mu$ and $ \Phi_L$  for input parameters
according to scenario $Snc_1$ with  $|\mu|=248$GeV.}
\label{fig:mass1_u=248_FU_FIL}
\end{figure}

\begin{figure} \centering
\begin{picture}(36.5,25)
\put(7,2){\includegraphics[width=70mm]{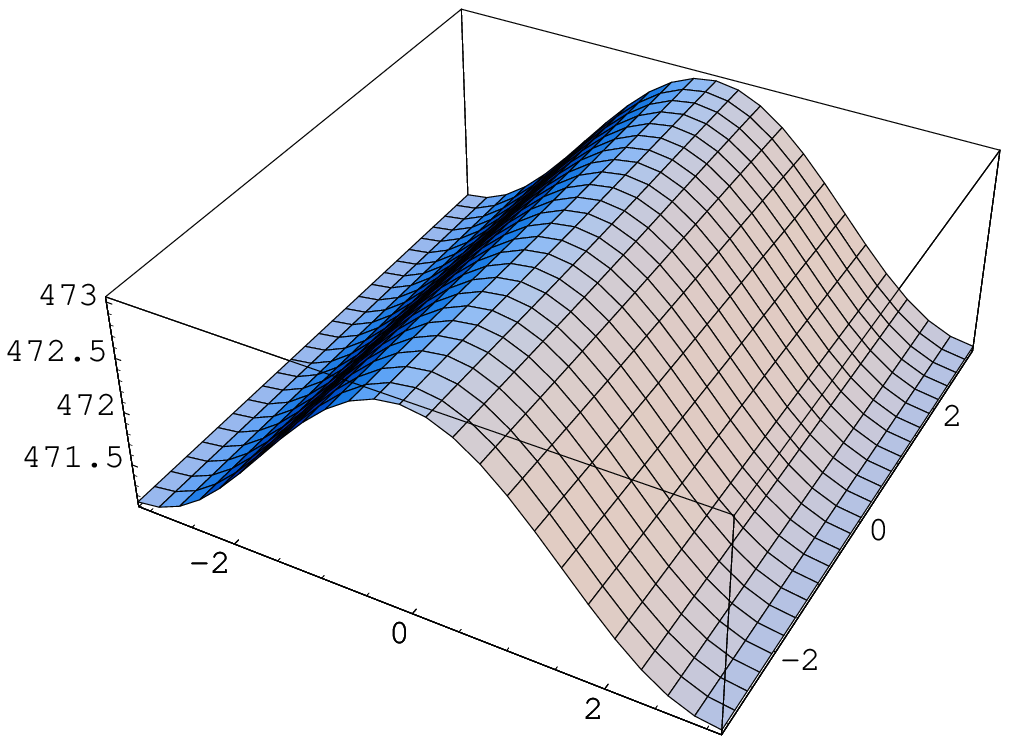}}
\put(13,3){\scriptsize{$\Phi_\mu \, {\rm (rad)}$}}
\put(30,6){\scriptsize{$\Phi_L \, {\rm (rad)}$}}
\put(1,10){\scriptsize{$m_{{\tilde \chi}^{0}_{1}} \, {\rm (GeV)}$}}
\end{picture}
\caption{Neutralino masse $m_{{\tilde \chi}^{0}_{2}}, $ as a
function of  $\Phi_\mu$ and $ \Phi_L$  for input parameters
according to scenario $Snc_1$ with  $|\mu|=248$GeV.}
\label{fig:mass2_u=248_FU_FIL}
\end{figure}

\subsection{The eigenvectors forming the matrix $V$} We have found
useful to finish this section with the computation of the matrix
$V.$ A more explicit form of this matrix will allow us to prove some
important relations in the next section.

The diagonalizing matrix $V$ can be obtained by computing the
eigenvectors corresponding to the eigenvalues given in Eq.
\eqref{eq:EIGV}. Indeed, by inserting a generic eigenvalue
$m_{{\tilde \chi}^{0}_{j}},$ into  Eq.  \eqref{eq:FE-EVP} and
dividing each one of these  equations by $V_{1j}$, where it is
assumed that $V_{1j}\neq 0$, we get \beqa \label{eq:Blur} \nonumber
H_{12}\,\frac{V_{2j}}{V_{1j}}+H_{13}\,\frac{V_{3j}}{V_{1j}}+H_{14}\,\frac{V_{4j}}{V_{1j}}-m_{{\tilde
\chi}^{0}_{j}}^2,
&=& -H_{11},\nonumber \\
\nonumber (H_{22}-m_{{\tilde
\chi}^{0}_{j}}^2)\,\frac{V_{2j}}{V_{1j}}+H_{23}\,\frac{V_{3j}}{V_{1j}}+H_{24}\,\frac{V_{4j}}{V_{1j}}
&=& -H_{21},\nonumber \\
\nonumber H_{32}\,\frac{V_{2j}}{V_{1j}}+ (H_{33}-m_{{\tilde
\chi}^{0}_{j}}^2)\,\frac{V_{3j}}{V_{1j}}+H_{34}\,\frac{V_{4j}}{V_{1j}}
&=& -H_{31}, \nonumber \\
\nonumber
H_{42}\,\frac{V_{2j}}{V_{1j}}+H_{43}\,\frac{V_{3j}}{V_{1j}}+
(H_{44}-m_{{\tilde \chi}^{0}_{j}}^2) \,\frac{V_{4j}}{V_{1j}} &=&
-H_{41}. \\ \eeqa
Solving this system of equations, and taking into account the
relation
\beqa \label{eq:Nabla} |V_{1j}|^2+ |V_{2j}|^2+
|V_{3j}|^2+|V_{4j}|^2=1, \eeqa
it yields  the $V_{ij}$ matrix's component \be \label{eq:EVEVij}
V_{ij} = {\Delta_{ij} \over \Delta_{1j}} \, {| \Delta_{1j} | \, e^{i
\theta_j}\over  \sqrt{| \Delta_{1j} |^2 + | \Delta_{2j} |^2 | +
|\Delta_{3j} |^2 | + |\Delta_{4j} |^2}}, \ee when i=1,\ldots,4.
Here, the $ \theta_j $'s are arbitrary phases, related to the CP
eigenphases, which will be fixed by the requirement that $V$
satisfies Eq. \eqref{eq:VMV}, as we will see in the next section,
\be \label{eq:delta1j} \Delta_{1j} = \begin{vmatrix} H_{22} -
m_{{\tilde \chi}^{0}_{j}}^2  & H_{23}& H_{24}\cr H_{32} & H_{33} -
m_{{\tilde \chi}^{0}_{j}}^2 & H_{34}\cr H_{42} &H_{43} &H_{44} -
m_{{\tilde \chi}^{0}_{j}}^2  \cr
\end{vmatrix}
\ee and $\Delta_{ij}, \, i=2,3,4,$ is formed from $\Delta_{1j}$ by
substituting  the $(i-1)$th  column by $\bigl(
\begin{smallmatrix}- H_{21} \cr - H_{31} \cr - H_{41} \cr \end{smallmatrix}\bigr).$

\section{The neutralino projectors, pseudoprojectors and CP eigenphases}
\label{sec-projectors} To describe the neutralino observables we can
use the projector formalism \cite{key1,key2}. The neutralino
projector matrices can be defined as\cite{key13}
\be \label{eq:PJ} P_{j}=P^{\dag}_{j}=V \, E_{j}\,V^{-1}, \ee so that
\be \label{eq:PjVVC} {P_{j}}_{\alpha \beta} = V_{\alpha j} \,
V^\ast_{\beta j}. \ee
These projectors satisfy the relations
\be \label{eq:proj-properties} P_{i}\,P_{j}=P_{j}\,\delta_{ij},\quad
Tr\,P_{j}=1,\quad\sum_{j=1}^{4}\,P_{j}=1, \ee
where $(i,j)=(1-4)$ describe the neutralino mass-eigenstate indices.
Notice that from  Eqs. \eqref{eq:MD2} and \eqref{eq:PJ} it is
possible to write  \be \label{eq:BMz}
N^{\dag}\,N=\sum_{j=1}^{4}\,m_{{\tilde \chi}^{0}_{j}}^{2}\,P_{j}.
\ee

As in the case of the study of the neutralino projector formalism
for complex supersymmetry parameters
 based on the MSSM \cite{key2}, here only the projectors are not
sufficient to describe the physical observables. For a complete
description of  physical observables it is also necessary to know
the so-called pseudoprojector matrices and  CP eigenphases. In the
following we implement a method, based on the explicit knowledge of
the diagonalizing  matrix $V$ to obtain these quantities and
demonstrate some of their properties.

\subsection{Reduced projectors}

By inserting \eqref{eq:EVEVij} into \eqref{eq:PjVVC},
 we get \be \label{eq:CPjalpha} P_{j \alpha
\beta} ={ p_{j \alpha} p^\ast_{j \beta} \over 1 + |p_{j2}|^2 +
|p_{j3}|^2 + |p_{j4}|^2},\ee where we define the reduced projectors
\be \label{eq:redu-proj} p_{j \alpha} \equiv {\Delta^\ast_{\alpha j}
\over \Delta^\ast_{1j}}.  \ee Notice that the expression given in
\eqref{eq:redu-proj} is a new version of the reduced projector
formula \cite{key2}. Indeed, from this last equation, it is clear
that $ p_{j 1}=1.$ Moreover, from Eq. \eqref{eq:CPjalpha} we deduce
\be \label{eq:Pj11}
 P_{j11} ={ 1  \over 1 + |p_{j2}|^2 +
|p_{j3}|^2 + |p_{j4}|^2}. \ee Thus, inserting this last result into
\eqref{eq:CPjalpha},  we prove the ansatz used in \cite{key2} \be
P_{j \alpha \beta} = P_{j11} \, p^\ast_{j \alpha} \, p_{j \beta}.
\label{clave-Jarlskog}\ee On the other hand,  using  Eqs.
\eqref{eq:redu-proj} and \eqref{eq:Pj11},
 we can write the matrix elements of the
diagonalizing matrix $V$ given in  Eq. \eqref{eq:EVEVij} in terms of
the reduced projectors, that is \be \label{eq:Vaj} V_{\alpha j} =
\sqrt{P_{j11} \over \eta_j}  p^\ast_{j \alpha}, \ee where $\eta_j
\equiv  e^{- 2 i \theta_j }$ stands for the CP eigenphases. As we
will see below,  this last equation allow us to express the L-R SUSY
parameters in terms of the reduced projectors and the eigenphases.

An useful property verified by the reduced projectors $p_{j \alpha}$
is
 \be \label{prop-reduced-proj} P_{j11} \,
\sum_{\beta=1}^4  p_{i \beta} \, p^\ast_{j \beta} = \delta_{ij}, \ee
which can be directly  deduced  from Eq. \eqref{eq:Vaj}, taking in
account the unitarirty of $V.$

Let us now to define other important matrices.   From Eq.
\eqref{eq:VMV},  we can write \be
 N = \sum_{j=1}^{4}\,m_{{\tilde \chi}^{0}_{j}}\,  V^{\ast}\,
 \,
E_{j} \,  V\dag = \sum_{j=1}^{4}\,m_{{\tilde \chi}^{0}_{j}}\, {\bar
P}_j, \label{eq:MVtVd} \ee where \be\label{eq:pseudo-proj} {\bar
P}_j \equiv V^{\ast}\,
 \,
E_{j} \,  V\dag \ee are the so-called pseudoprojectors \cite{key2}.
Using Eq. \eqref{eq:Vaj} and the definition  of $E_j,$ the matrix
elements of these pseudoprojectors can  easily be written in the
form \be\label{eq:pseudoPab} {\bar P}_{j \alpha \beta} =
V^\ast_{\alpha j} V^\ast_{\beta j}= P_{j11} \, p_{j \alpha} \, p_{j
\beta} \, \eta_j. \ee From this last equation it is clear that
${\bar P}_j $ is a symmetric matrix, that is  ${\bar P}_{j}^T =
{\bar P}_{j}. $

Using  Eq. \eqref{eq:pseudoPab}, taking again  into account the
unitarity of $V,$  and  the definition \eqref{eq:PjVVC},  we have
\begin{eqnarray*} ({\bar P}_j^\ast {\bar P}_k )_{\alpha \beta} &=&
\sum_{\rho=1}^4 {\bar P}^\ast_{j \alpha \rho} {\bar P}_{k \rho
\beta} \\ &=& \sum_{\rho=1}^4 V_{\alpha j} V_{\rho j} V^\ast_{\rho
k} V^\ast_{\beta k} = \delta_{jk}
  V_{\alpha j}  V^\ast_{\beta j} =
\delta_{jk} P_{j \alpha \beta},\end{eqnarray*} that is, the
pseudoprojectors satisfy \be {\bar P}_j^\ast {\bar P}_k  =
\delta_{jk} P_j. \ee In the same way, we can show that \be {\bar
P}_j^\ast \, P_k = P_k^t \, {\bar P}_j = \delta_{jk} \,{\bar P}_j.
\ee

As we have mentioned in the previous section, the eigenphases
$\eta_j $ must be chosen in such a way that the diagonalizing matrix
$V$ satisfies Eq. \eqref{eq:VMV} or equivalently Eq.
\eqref{eq:MVtVd}. Inserting  Eq. \eqref{eq:pseudoPab} into Eq.
\eqref{eq:MVtVd} and using the property Eq.
\eqref{prop-reduced-proj} we get \be \label{eq:etaj} \eta_j \,
m_{{\tilde \chi}^{0}_{j}} = \sum_{\alpha =1}^4  N_{\alpha \beta} \,
{p^\ast_{j \alpha} \over p_{j\beta}}= \sum_{\alpha =1}^4 N_{\alpha
\beta} \, {\Delta_{\alpha j} \over \Delta_{\beta j}^\ast}. \ee This
last Equation represents, for fixed $j,$ four equivalent relations
serving to determine the fundamental parameters of the model, namely
$M_L, \mu, M_R,  M_V $ and $\tan \theta_k, $ in terms of the reduced
projectors, the physical neutralino masses, the eigenphases and the
L-R SUSY  coupling constants. We notice that, starting from Eq.
\eqref{eq:VMV} and using Eq. \eqref{eq:Vaj}, a more symmetric
structure for the eigenphases $\eta_j$ can be reached, that is
 \be
\eta_j \, m_{{\tilde \chi}^{0}_{j}} = P_{j11} \, \sum_{\alpha, \beta
=1}^4 p^\ast_{j \alpha} \,  N_{\alpha \beta} \, p^\ast_{j \beta}.
 \ee
This  relation can also be constructed directly from the more
fundamental Eq. \eqref{eq:etaj},  by  means of  property Eq.
\eqref{prop-reduced-proj}.

\subsection{Explicit form of the reduced projectors}
According to Eq. \eqref{eq:redu-proj}, to obtain the explicit form
of the reduced projectors in terms of the fundamental parameters of
the theory  only we need  to know the explicit form of quantities $
\Delta_{\alpha j}^\ast.$ For fixed $j,$ they are given by
\beqa\nonumber \Delta^\ast_{1 j} &=& - 4 \kappa^2 (1 + 4 \kappa^2 )
M^4 (m_{{\tilde \chi}^{0}_{j}}^2 -
 4 |\mu|^2 \sin^2( 2 \theta_k))\\ \nonumber &+&
 M^2 (m_{{\tilde \chi}^{0}_{j}}^2 -
 4 |\mu|^2 ) \bigl[ m_{{\tilde \chi}^{0}_{j}}^2 - M_{RV}^2  \\\nonumber &+&  8
 \kappa^2 m_{{\tilde \chi}^{0}_{j}}^2 +
 16 \kappa^2 M_{RV} |\mu| \cos\Phi_\mu \sin(2\theta_k)  \bigr]
\\  &-&( m_{{\tilde \chi}^{0}_{j}}^2 - M_{RV}^2 ) ( m_{{\tilde \chi}^{0}_{j}}^2 - 4
|\mu|^2)^2, \label{eq:cdelta1j}\eeqa \beqa \nonumber \Delta^\ast_{2
j} &=&  - 2 \kappa (1 + 4 \kappa^2) M^4 [m_{{\tilde \chi}^{0}_{j}}^2
- 4 |\mu|^2 \sin^2 (2 \theta_k)] \\ \nonumber&+& 2 \kappa M^2
[m_{{\tilde \chi}^{0}_{j}}^2 - 4 |\mu|^2] \\\nonumber &\times&
\bigl\{ m_{{\tilde \chi}^{0}_{j}}^2 + M_{RV} |M_L| e^{-i \Phi_L} + 2
|\mu| \sin(2\theta_k)\\ &\times& [ |M_L| e^{-i (\Phi_L + \Phi_\mu)}
+
 M_{RV} e^{i  \Phi_\mu} ] \bigr\},\label{eq:cdelta2j}
 \eeqa
\beqa \nonumber \Delta^\ast_{3 j} &=&  2 M^3 \bigl\{ |\mu|
\cos\theta_k \cos(2 \theta_k) \bigl[4 \kappa^2 M_{RV} |M_L| e^{i
(\Phi_\mu - \Phi_L)}
\\ \nonumber &-& (m_{{\tilde \chi}^{0}_{j}}^2 (1 + 4 \kappa^2 ) -
M_{RV}^2) e^{i \Phi_\mu}\bigr] +  2 \kappa^2 \sin\theta_k \\
\nonumber  &\times& (m_{{\tilde \chi}^{0}_{j}}^2 - 8 |\mu|^2 \cos^2
\theta_k ) (M_{RV} - |M_L| e^{- i \Phi_L }) \bigr\}
\\ \nonumber &+& M (m_{{\tilde \chi}^{0}_{j}}^2 - 4
|\mu|^2) (m_{{\tilde \chi}^{0}_{j}}^2 - M_{RV}^2) \\
&\times& (|M_L| e^{-i \Phi_L} \sin\theta_k + 2 |\mu| e^{i \Phi_\mu}
\cos\theta_k) \label{eq:cdelta3j} \eeqa and
 \beqa \nonumber \Delta^\ast_{4 j} &=& 2 M^3 \bigl\{
|\mu| \sin\theta_k \cos(2 \theta_k) \bigl[4 \kappa^2 M_{RV} |M_L|
e^{i (\Phi_\mu - \Phi_L)} \\ \nonumber &-& (m_{{\tilde
\chi}^{0}_{j}}^2 (1 + 4 \kappa^2 ) - M_{RV}^2) e^{i \Phi_\mu}\bigr]
-  2 \kappa^2 \cos\theta_k \\\nonumber  &\times& (m_{{\tilde
\chi}^{0}_{j}}^2 - 8 |\mu|^2 \sin^2 \theta_k ) (M_{RV} - |M_L| e^{-
i \Phi_L }) \bigr\}
\\ \nonumber &-& M (m_{{\tilde \chi}^{0}_{j}}^2 - 4
|\mu|^2) (m_{{\tilde \chi}^{0}_{j}}^2 - M_{RV}^2) \\
&\times& (|M_L| e^{-i \Phi_L} \cos\theta_k + 2 |\mu| e^{i \Phi_\mu}
\sin\theta_k) \label{eq:cdelta4j}.\eeqa

The formulas (\ref{eq:cdelta1j}-\ref{eq:cdelta4j}) allow us to
express, through the reduced projectors, all the essential
quantities of the model in terms of the original parameters.

\subsection{Consistence with the Jarlskog's formula}
\label{sec-Jarlskog-equivalence} Using  the projector properties
\eqref{eq:proj-properties} and some ones associated to the
coefficients of the characteristic polynomial \eqref{eq:CEQ}, we can
write  the projectors $P_j$ in terms  of the neutralino masses and
the $H$ matrix, in  the Jarlskog's forme \cite{key7}: \beqa
 P_{1} &=&
\frac{(m_{{\tilde \chi}^{0}_{4}}^2-H)(m_{{\tilde
\chi}^{0}_{3}}^2-H)(m_{{\tilde \chi}^{0}_{2}}^2-H)}{(m_{{\tilde
\chi}^{0}_{4}}^2-m_{{\tilde \chi}^{0}_{1}}^2)(m_{{\tilde
\chi}^{0}_{3}}^2-m_{{\tilde \chi}^{0}_{1}}^2)
(m_{{\tilde \chi}^{0}_{2}}^2-m_{{\tilde \chi}^{0}_{1}}^2)}, \nonumber \\
P_{2} &=& \frac{(m_{{\tilde \chi}^{0}_{1}}^2-H)(m_{{\tilde
\chi}^{0}_{4}}^2-H)(m_{{\tilde \chi}^{0}_{3}}^2-H)}{(m_{{\tilde
\chi}^{0}_{1}}^2-m_{{\tilde \chi}^{0}_{2}}^2)(m_{{\tilde
\chi}^{0}_{4}}^2-m_{{\tilde \chi}^{0}_{2}}^2)
(m_{{\tilde \chi}^{0}_{3}}^2-m_{{\tilde \chi}^{0}_{2}}^2)}, \nonumber \\
P_{3} &=& \frac{(m_{{\tilde \chi}^{0}_{2}}^2-H)(m_{{\tilde
\chi}^{0}_{1}}^2-H)(m_{{\tilde \chi}^{0}_{4}}^2-H)}{(m_{{\tilde
\chi}^{0}_{2}}^2-m_{{\tilde \chi}^{0}_{3}}^2)(m_{{\tilde
\chi}^{0}_{1}}^2-m_{{\tilde \chi}^{0}_{3}}^2)
(m_{{\tilde \chi}^{0}_{4}}^2-m_{{\tilde \chi}^{0}_{3}}^2)}, \nonumber \\
P_{4} &=& \frac{(m_{{\tilde \chi}^{0}_{3}}^2-H)(m_{{\tilde
\chi}^{0}_{2}}^2-H)(m_{{\tilde \chi}^{0}_{1}}^2-H)}{(m_{{\tilde
\chi}^{0}_{3}}^2-m_{{\tilde \chi}^{0}_{4}}^2)(m_{{\tilde
\chi}^{0}_{2}}^2-m_{{\tilde \chi}^{0}_{4}}^2) (m_{{\tilde
\chi}^{0}_{1}}^2-m_{{\tilde \chi}^{0}_{4}}^2)}. \nonumber
\\ \label{Jarlskog-uno} \eeqa A more useful expression for these projectors is obtained if
we define \be P_j = {{\tilde P}_j \over {\tilde \Delta}_j},
\label{eq:ptilde-grande} \ee where \be {\tilde \Delta}_j = - 3
m_{{\tilde \chi}^{0}_{j}}^8 + 2 a \, m_{{\tilde \chi}^{0}_{j}}^6 - b
\, m_{{\tilde \chi}^{0}_{j}}^4 + d. \ee Indeed, by performing some
algebraic manipulations we get \beqa \label{eq:ptildej} {\tilde
P}_{j \alpha \beta} &=&- m_{{\tilde \chi}^{0}_{j}}^6 \, H_{\alpha
\beta} +
m_{{\tilde \chi}^{0}_{j}}^4 \, (a \, H_{\alpha \beta} - H^2_{\alpha \beta} ) \nonumber \\
&+& m_{{\tilde \chi}^{0}_{j}}^2 \, (a \,  H^2_{\alpha \beta} - b \,
H_{\alpha \beta} - H^3_{\alpha \beta}) + d \, \delta_{\alpha \beta}.
\eeqa

Now, combining  Eqs. \eqref{clave-Jarlskog} and
\eqref{eq:ptilde-grande}, we  deduce the expression
\be\label{eq:redu-proj-standar}
 p_{j \alpha} = { P_{j 1 \alpha} \over  P_{j11} } = { {\tilde P}_{j 1 \alpha} \over  {\tilde P}_{j11} }, \ee
 which can also be considered as a definition for the reduced
 projectors.

Equations \eqref{eq:redu-proj} and \eqref{eq:redu-proj-standar} are
equivalent expressions for the reduced projectors  when we
substitute into them the exact analytical masses $m_{{\tilde
\chi}^{0}_{j}}$ given  in \eqref{eq:EIGV}. Thus, combining these
equations and comparing the expressions
(\ref{eq:cdelta2j}-\ref{eq:cdelta4j}) with the corresponding
${\tilde P}_{j 1 \beta}, \beta=2,3,4,$ computed from Eq.
\eqref{eq:ptildej}, we can show that \be \label{eq:ptilde-delta}
{\tilde P}_{j 1 \alpha} = m_{{\tilde \chi}^{0}_{j}}^2 \,
 \Delta^\ast_{\alpha j}, \quad \forall \, \alpha=1,\ldots,4,  \ee
with \beqa\nonumber  {\tilde P}_{j 1 1} &=& M^4 [m_{{\tilde
\chi}^{0}_{j}}^2 -
 4 |\mu|^2 \sin^2( 2 \theta_k)] [m_{{\tilde \chi}^{0}_{j}}^2
 (1+ 4 \kappa^2 ) \\\nonumber &-& 16 \kappa^4 |M_L|^2 - M_{RV}^2 - 8 \kappa^2
 |M_L| M_{RV} \cos\Phi_L ]\\\nonumber &-& M^2
 ( m_{{\tilde \chi}^{0}_{j}}^2 - 4 |\mu|^2) \bigl\{  2 |\mu| |M_L|
 \sin(2\theta_k) \\\nonumber &\times&
\bigl[ 2 (m_{{\tilde \chi}^{0}_{j}}^2 - M_{RV}^2 ) \cos(\Phi_L +
\Phi_\mu) \\\nonumber &-& 8 |M_L| M_{RV} \kappa^2 \cos\Phi_\mu
\bigr]
\\ \nonumber&+& m_{{\tilde \chi}^{0}_{j}}^2 \bigl[(m_{{\tilde \chi}^{0}_{j}}^2 - M_{RV}^2) - 8 \kappa^2
|M_L|^2\bigr] \bigr\} \\ &-& |M_L|^2 (m_{{\tilde \chi}^{0}_{j}}^2 -
M_{RV}^2) ( m_{{\tilde \chi}^{0}_{j}}^2 - 4 |\mu|^2)^2.
\label{eq:ptilde11}\eeqa Indeed, Eqs. \eqref{eq:ptilde-delta}, for
$\alpha=1,2,3,$ constitute an identity whereas for $\alpha=1$ it
constitutes an useful equivalence, as we will show in the next
section.

\section{General disentangle formula of  $M_{L}$ in terms of the
eigenphases} \label{sec-genral-ml-eigenphases}
 From Eq. \eqref{eq:etaj},  choosing $\beta=1$ and using \eqref{eq:Imatrix}, we get
\be \eta_j  m_{{\tilde \chi}^{0}_{j}} = M_{L} - M  {\sin \theta_k
\Delta_{3j} - \cos\theta_k \Delta_{4j} \over \Delta_{1j}^\ast}.
\label{eq:etaj-mj} \ee Inserting \eqref{eq:cdelta3j}  and
\eqref{eq:cdelta4j} into \eqref{eq:etaj-mj}  and solving a  linear
algebraic equation for $M_{L},$ we get \beqa \nonumber M_{L} &=& {
\Delta_{1j}^\ast m_{{\tilde \chi}^{0}_{j}} \over {\cal D}_j} \eta_j
+  {2 M^2 \over
{\cal D}_j} \bigl\{|\mu| e^{-i \Phi_\mu} \sin(2 \theta_k) \\
\nonumber &\times& (m_{{\tilde \chi}^{0}_{j}}^2 -M_{RV}^2)
(m_{{\tilde \chi}^{0}_{j}}^2 -4 |\mu|^2 ) \\ &+& 2 \kappa^2 M^2
M_{RV} [m_{{\tilde
\chi}^{0}_{j}}^2 -4 |\mu|^2 \sin^2(2 \theta_k) ] \bigr\} \nonumber \\
&=& A_j \, \eta_j + B_j, \label{genMLformula} \eeqa where \beqa
\nonumber A_j &=& { \Delta_{1j}^\ast m_{{\tilde \chi}^{0}_{j}} \over
{\cal D}_j}= - {m_{{\tilde \chi}^{0}_{j}}\over {\cal D}_j } \bigl\{
(m_{{\tilde \chi}^{0}_{j}}^2 -M_{RV}^2) (m_{{\tilde \chi}^{0}_{j}}^2
-4 |\mu|^2 )^2 \\ \nonumber &+& 4 \kappa^2 (1+4 \kappa^2) M^4
[m_{{\tilde \chi}^{0}_{j}}^2 -4 |\mu|^2 \sin^2(2 \theta_k) ]
\\\nonumber &-& M^2
(m_{{\tilde \chi}^{0}_{j}}^2 -4 |\mu|^2 ) [ 8 \kappa^2 m_{{\tilde
\chi}^{0}_{j}}^2  + (m_{{\tilde \chi}^{0}_{j}}^2 - M_{RV}^2 ) \\ &+&
16 \kappa^2 |\mu| M_{RV} \cos\Phi_\mu \sin(2\theta_k)] \bigr\}
\label{eq:AJ}, \eeqa \beqa \nonumber B_j&=& {2 M^2 \over {\cal D}_j}
\bigl\{|\mu| e^{-i \Phi_\mu} \sin(2 \theta_k)\\\nonumber &\times&
(m_{{\tilde \chi}^{0}_{j}}^2 -M_{RV}^2) (m_{{\tilde \chi}^{0}_{j}}^2
-4 |\mu|^2 ) \\ &+& 2 \kappa^2 M^2 M_{RV} [m_{{\tilde
\chi}^{0}_{j}}^2 -4 |\mu|^2 \sin^2(2 \theta_k) ] \bigr\}
\label{eq:BJ} \eeqa and \beqa \nonumber {\cal D}_j &=& - \bigl\{
(m_{{\tilde \chi}^{0}_{j}}^2 - M_{RV}^2 )(m_{{\tilde
\chi}^{0}_{j}}^2 - 4 |\mu|^2)^2 - 8
\kappa^2 M^2 \\
\nonumber &\times&(m_{{\tilde \chi}^{0}_{j}}^2 - 4 |\mu|^2)
[m_{{\tilde \chi}^{0}_{j}}^2 + 2 |\mu| M_{RV} \cos\Phi_\mu \sin (2
\theta_k) ] \\ &+& 16 \kappa^4 M^4  [ m_{{\tilde \chi}^{0}_{j}}^2 -
4 |\mu|^2 \sin^2(2\theta_k)] \bigr\}. \label{eq:cal-dej} \eeqa
\begin{table}
\begin{center}
\begin{tabular}{c c c c c c }\cline{1-6}\\
Scenario & $|\mu| $\; &  $M_R \; $ & $ M_V \; $ &  $ \tan\theta_k  $ \; &  $ \eta_j $   \\
\\\cline{1-6} \\
$Sc_{3a} $ &  248 & 500 & 50  &   30  & \parbox[c]{0.5cm} {1 \\ -1 } \\ \\  \hline \\
$Sc_{3b} $ &  248 & 500 & 500  &   30  & \parbox[c]{0.5cm} {1 \\ -1 } \\ \\  \hline \\
$Sc_{3c} $ &  500 & 500 & 50  &   30  & \parbox[c]{0.5cm} {1 \\ -1 } \\ \\  \hline \\
$Sc_{3d} $ &  248 & 500 & 50  &   \parbox[c]{0.5cm} {10\\50}  & \parbox[c]{0.5cm} {1 \\ -1 } \\ \\  \hline \\
\end{tabular}
\end{center}
\caption{Input parameters for scenarios $Sc_{3a}, \ldots, Sc_{3d}.$
All mass quantities are in GeV.} \label{tab:tablaML1}
\end{table}

Equation \eqref{genMLformula} allow us to determinate the behavior
of  $|M_L|$ and $\Phi_L$ in terms of the eigenphases $\eta_j$ and
the physical masses $m_{{\tilde \chi}^{0}_{j}},$ when the rest of
fundamental parameters are fixed. We notice that this  equation has
been obtained without use  the Jarlskog's projector formula
\eqref{Jarlskog-uno} or its equivalent \eqref{eq:ptildej}. The
method used to obtain it is direct and it is based essentially on
the fact that $\Delta_{1 j}$ is  independent of $|M_L|$ and of
$\Phi_L.$

\begin{figure} \centering
\begin{picture}(31.5,21)
\put(1,2){\includegraphics[width=70mm]{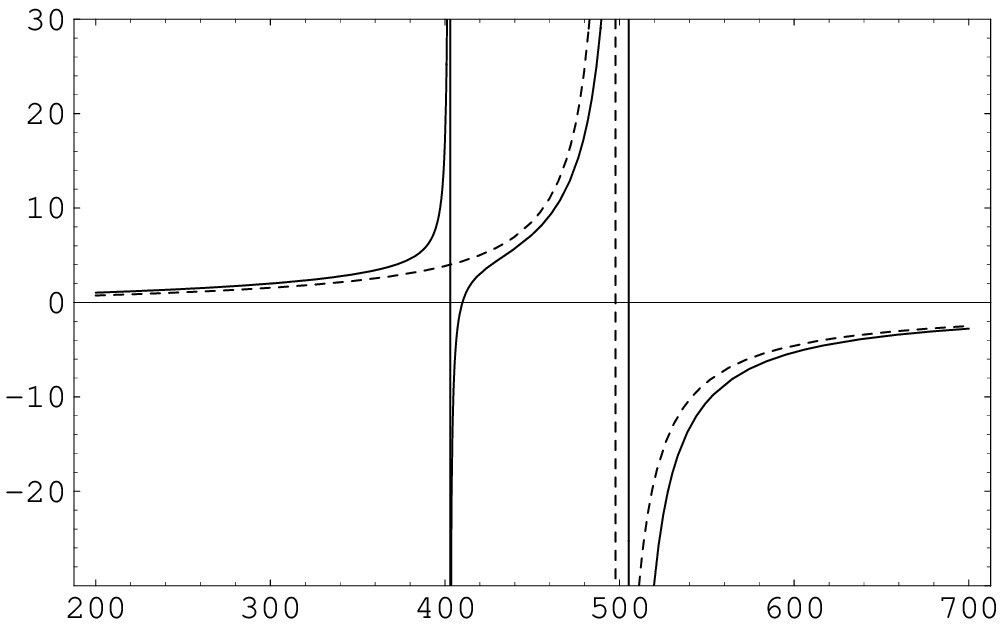}}
\put(15.75,0.7){\scriptsize{$m_{{\tilde \chi}^{0}_{j}}$}}
\put(1.05,19.25){\scriptsize{$(|M_L| - m_{{\tilde
\chi}^{0}_{j}})/2$}}
\end{picture}
\caption{Graph of  $|M_L|,$ using the general formula
\eqref{genMLformula}, for inputs of scenario $Sc_{3a},$ with
$\eta_j=1$ (solid line) and $\eta_j=-1$ (dashed line), as  a
function of the physical neutralino masses $m_{{\tilde
\chi}^{0}_{i}}.$ } \label{fig:MLgenmj1}
\end{figure}

\begin{figure} \centering
\begin{picture}(31.5,21)
\put(1,2){\includegraphics[width=70mm]{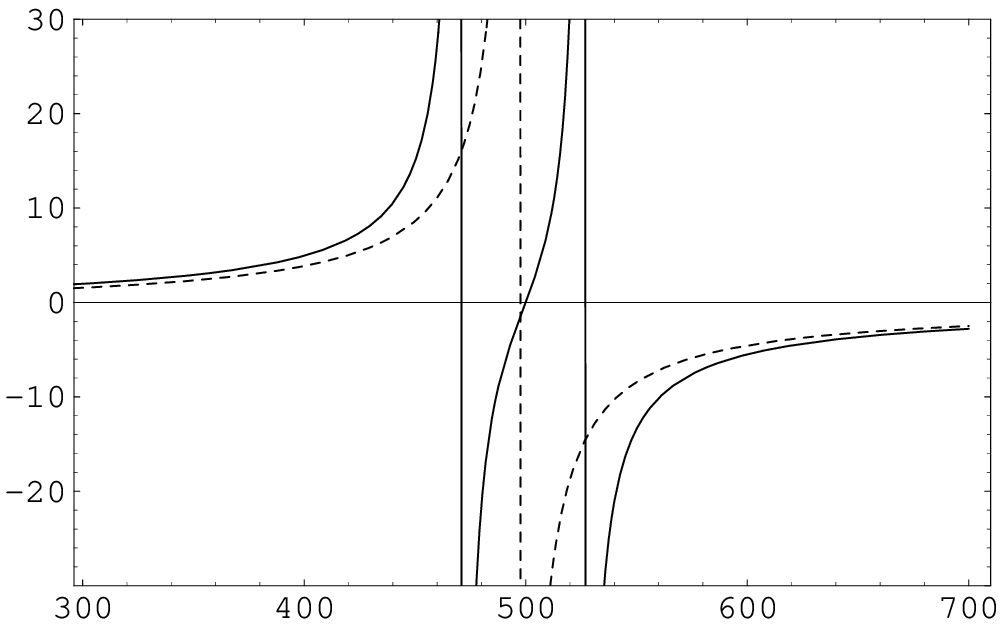}}
\put(15.75,0.7){\scriptsize{$m_{{\tilde \chi}^{0}_{j}}$}}
\put(1.05,19.25){\scriptsize{$(|M_L| - m_{{\tilde
\chi}^{0}_{j}})/2$}}
\end{picture}
\caption{Graph of  $|M_L|,$ using the general formula
\eqref{genMLformula}, for inputs of scenario $Sc_{3b},$ with
$\eta_j=1$ (solid line) and $\eta_j=-1$ (dashed line), as  a
function of the physical neutralino masses $m_{{\tilde
\chi}^{0}_{i}}.$ } \label{fig:MLgenmj1c}
\end{figure}

\begin{figure} \centering
\begin{picture}(31.5,21)
\put(1,2){\includegraphics[width=70mm]{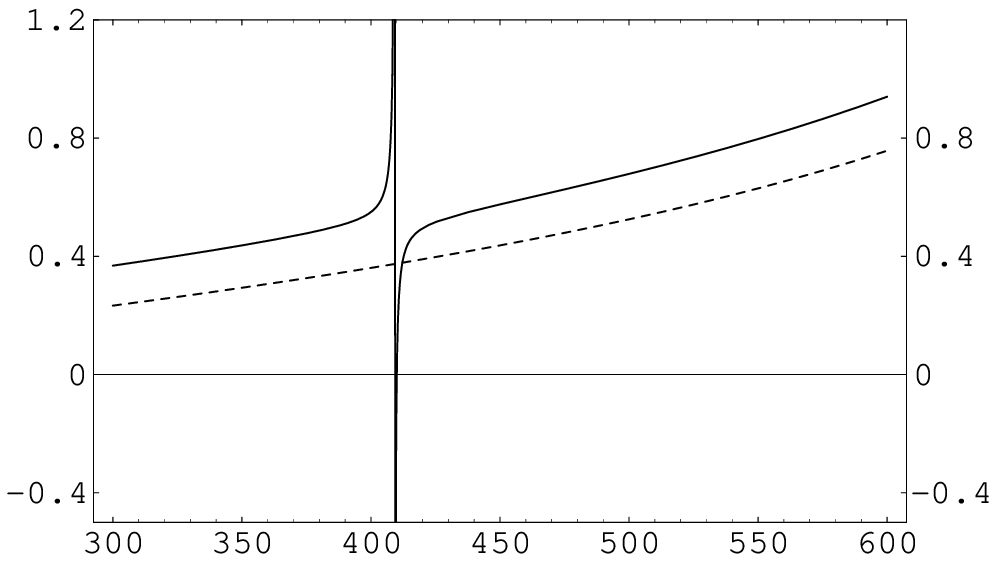}}
\put(15.75,0.7){\scriptsize{$m_{{\tilde \chi}^{0}_{j}}$}}
\put(1.05,19.25){\scriptsize{$(|M_L| - m_{{\tilde
\chi}^{0}_{j}})/2$}}
\end{picture}
\caption{Graph of  $|M_L|,$ using the general formula
\eqref{genMLformula}, for inputs of scenario $Sc_{3c},$ with
$\eta_j=1$ (solid line) and $\eta_j=-1$ (dashed line), as  a
function of the physical neutralino masses $m_{{\tilde
\chi}^{0}_{i}}.$ } \label{fig:MLgenmj1b}
\end{figure}

\begin{figure} \centering
\begin{picture}(31.5,21)
\put(1,2){\includegraphics[width=70mm]{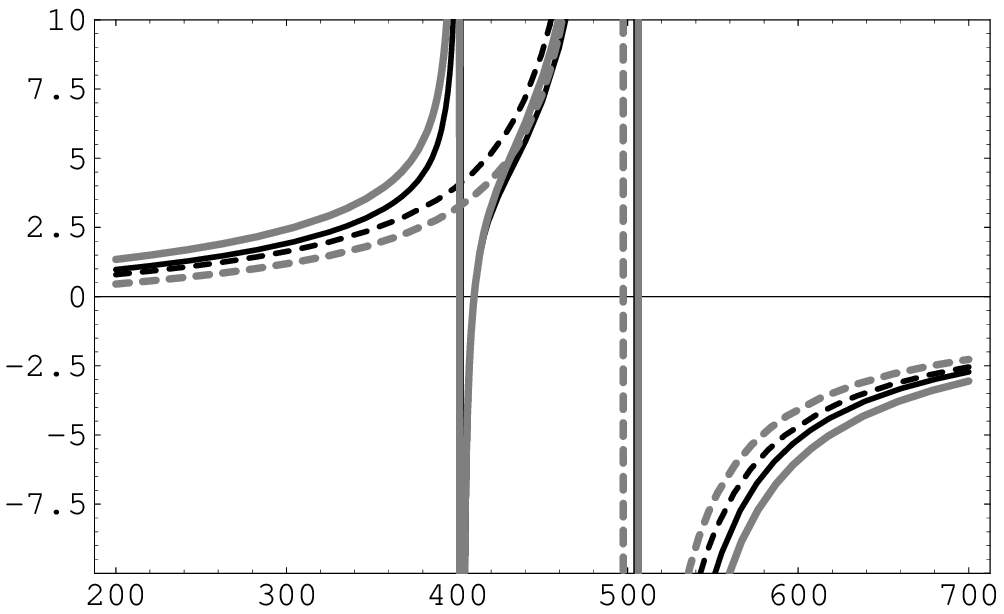}}
\put(15.75,0.7){\scriptsize{$m_{{\tilde \chi}^{0}_{j}}$}}
\put(1.05,19.25){\scriptsize{$(|M_L| - m_{{\tilde
\chi}^{0}_{j}})/2$}}
\end{picture}
\caption{Graph of  $|M_L|$ as a function of the physical neutralino
masses $m_{{\tilde \chi}^{0}_{i}},$ using the general formula
\eqref{genMLformula}, for inputs of scenario $Sc_{3d}.$ The curves
are: $\tan\theta_k=10, \eta_j=1$ (light solid); $\tan\theta_k=10,
\eta_j=-1$ (light dashed); $\tan\theta_k=50, \eta_j=1$ (heavy solid)
and $\tan\theta_k=50, \eta_j=-1$ (heavy dashed).}
\label{fig:MLgenmj1d}
\end{figure}

Figure \ref{fig:MLgenmj1} shows the behavior of $|M_L|$ as a
function of the neutralino  masses  $m_{{\tilde \chi}^{0}_{j }},$
for input parameters of the CP-conserving  scenario $Sc_{3a},$ given
in Tab. \ref{tab:tablaML1} (as before, we assume $g_L=g_R=g_V=0.65$,
and $k_u=92.75$). We observe that for small values of the neutralino
masses, i.e., for masses of order $200\, {\rm GeV}$ approximately,
the size of $|M_L|$ becomes the same in both cases, the scenario
$Sc_{3a}$  with $\eta_j=1$ and the scenario $Sc_{3a}$ with
$\eta_j=-1.$ Let us now  consider the scenario  $Sc_{3b},$ which is
the same as the scenario $Sc_{3a},$  except for the value of $M_V$
which have been increased from  $50$ GeV to $500$ GeV. In this case,
the common value of  $|M_L|$  in  both scenarios, i.e., $Sc_{3b}$
with $\eta=\pm 1,$ is found to be $|M_L| \approx 300$ GeV, in the
region of small physical neutralino masses of the order of $300$
GeV, as we can see from Fig \ref{fig:MLgenmj1c}.  Figure
\ref{fig:MLgenmj1b} shows the behavior of $|M_L|$ as a function of
the neutralino masses $m_{{\tilde \chi}^{0}_{j }},$ for input
parameters of the CP-conserving scenario $Sc_{3c},$ given in Tab.
\ref{tab:tablaML1}. This scenario differs from the scenario
$Sc_{3a}$ in the value of $\mu$ which have now been taken $|\mu|=
500$ GeV. We observe that the curves corresponding to the input
parameters of  scenario $Sc_{3c}$ with different eigenphases values,
i.e., $\eta_j= \pm 1,$ intersect when
 $m_{{\tilde \chi}^{0}_{j}}\approx 412.17 $GeV, giving the common  value
 of  $|M_L|\approx 412.93$ GeV, which is bigger than the corresponding common values of $|M_L|$ computed in the
 previous scenarios. On the other hand,  if we compare the curves representing the behavior of $|M_L|$ as a function
 of the physical neutralino masses for different values of the
 parameter $\tan\theta_k,$ we don't observe  important
 differences between them when the values of this last parameter is
 chosen in the range 30-50, however, if we compare the mentioned
 curves for values of $\tan\theta_k$  chosen, for instance, in the range 10-30 or 10-50, we observe that
 for small neutralino masses $m_{{\tilde
\chi}^{0}_{i}},$  the values of $|M_L|$ approach from the right to
the given value of  $m_{{\tilde \chi}^{0}_{i}},$ and this approach
is more significative for big values of $\tan\theta_k$ than for
small ones when $\eta_j =1$ and viceversa, this approach is more
significative for small values of $\tan\theta_k$ than for big ones
when $\eta_j = -1.$ This means that the value of the light
neutralino mass which provides the value of $|M_L|$ which is
independent of the eigenphases $\eta_j = \pm 1,$ increases when
$\tan\theta_k$ augments. The above mentioned  behavior of the
parameters is verified by seen the plots in Fig.
\ref{fig:MLgenmj1d}, where we plot $|M_L|$ versus $m_{{\tilde
\chi}^{0}_{i}},$  for inputs of scenario $Sc_{3d}$ with
$\tan\theta_k=10$ and $\tan\theta_k =50.$

\subsection{An alternative way to obtain $|M_L|$} \label{sec-FUND-PARAMETERS}
 When $\alpha=1,$ Eq. \eqref{eq:ptilde-delta} combined with  Eqs. \eqref{eq:cdelta1j} and
 \eqref{eq:ptilde11}, allow us to express the norm of $M_L$ in terms  of the rest
 of the fundamental parameters $\Phi_L, |\mu|, \Phi_\mu, \tan\, \theta_k $ and $M_{RV}$ and the physical masses
 $m_{{\tilde \chi}^{0}_{j}}$. Indeed, inserting
\eqref{eq:cdelta1j} and \eqref{eq:ptilde11} into
\eqref{eq:ptilde-delta} and solving a quadratic algebraic equation
for $|M_L|,$ we get \be \label{eq:normeML} |M_L| ={ - {\cal B}_j \pm
\sqrt{{\cal B}_j^2 - 4 {\cal D}_j ({\cal C}_j - m_{{\tilde
\chi}^{0}_{j}}^2 \Delta^\ast_{1 j} )} \over 2 {\cal D}_j}, \ee where
\beqa \nonumber {\cal B}_j &=& - 4 M^2 \bigl\{ |\mu|(m_{{\tilde
\chi}^{0}_{j}}^2 - M_{RV}^2 )(m_{{\tilde \chi}^{0}_{j}}^2 - 4
|\mu|^2) \\ \nonumber &\times&
\cos(\Phi_L + \Phi_\mu) \sin(2 \theta_k) + 2\kappa^2 M^2 M_{RV} \\
&\times& [ m_{{\tilde \chi}^{0}_{j}}^2 - 4 |\mu|^2
\sin^2(2\theta_k)] \cos\Phi_L \bigr\} ,\eeqa  \beqa \nonumber {\cal
C}_j &=& - M^2\, m_{{\tilde \chi}^{0}_{j}}^2 ( m_{{\tilde
\chi}^{0}_{j}}^2 - M_{RV}^2) ( m_{{\tilde \chi}^{0}_{j}}^2 - 4
|\mu|^2) \\\nonumber  &+& M^4   [
m_{{\tilde \chi}^{0}_{j}}^2 (1 + 4 \kappa^2) - M_{RV}^2] \\
&\times& [ m_{{\tilde \chi}^{0}_{j}}^2 - 4 |\mu|^2
\sin^2(2\theta_k)]
  \eeqa and ${\cal D}_j$ is given in Eq. \eqref{eq:cal-dej}.

The formula for $|M_L|$ given in Eq.  \eqref{eq:normeML},
constitutes an alternative to the one given in Eq.
\eqref{genMLformula}, serving to study the behavior of $|M_L|$ as a
function of the phase $\Phi_L,$ the physical mass $m_{{\tilde
\chi}^{0}_{j}},$ and the rest of the fundamental parameters. For
instance, let us consider the possible CP-conserving scenario $Sc_4$
given in Tab. \ref{tab:tablaML2}.
\begin{table}
\begin{center}
\begin{tabular}{c c c c c c }\cline{1-6}\\
Scenario & $|\mu| $\; &  $M_R \; $ & $ M_V \; $ &  $ \tan\theta_k  $ \; &  $ \Phi_L $   \\
\\\cline{1-6} \\
$Sc_4 $ &  248 & 500 & 50  &   30  & \parbox[c]{0.5cm} {0 \\ $\pi$ } \\ \\  \hline \\
\end{tabular}
\end{center}
\caption{Input parameters for scenario $Sc_4.$ All mass quantities
are in GeV.} \label{tab:tablaML2}
\end{table} In this
case, the behavior of $|M_L|$ in terms of one of the physical mass
$m_{{\tilde \chi}^{0}_{j}},$ is shown in Figs. \ref{fig:MLvsmj1} and
\ref{fig:MLvsmj2}. It is clair that superposing Figures
\ref{fig:MLvsmj1} and \ref{fig:MLvsmj2}, we reconstruct Figure
\ref{fig:MLgenmj1}. Comparing these figures, we also observe that,
in the CP-conserving case, when $\Phi_\mu=0,$ the eigenphase values
$\eta_j=\pm 1$ correspond to the $M_L$ phase values $\Phi_L= \pm
\pi,$ respectively. The same considerations are valid when we take
$\Phi_\mu=\pi.$ That is,   in the CP-conserving case, when all the
parameters but $m_{{\tilde \chi}^{0}_{j}}$ are fixed, the choice of
the two different values  $\Phi_L= 0,\pi$ in Eq. \eqref{eq:normeML},
correspond to the choice of  the two possible values of the
eigenphases $\eta_j=1, -1,$  in  Eq. \eqref{genMLformula} .

\begin{figure} \centering
\begin{picture}(31.5,21)
\put(1,2){\includegraphics[width=70mm]{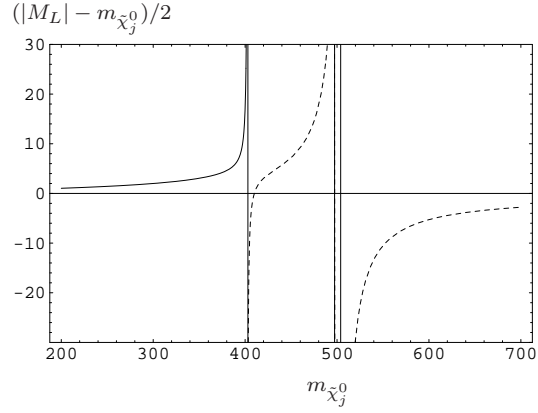}}
\put(15.75,0.7){\scriptsize{$m_{{\tilde \chi}^{0}_{j}}$}}
\put(1.05,19.25){\scriptsize{$(|M_L| - m_{{\tilde
\chi}^{0}_{j}})/2$}}
\end{picture}
\caption{Graph of $|M_L|$ as a function of the physical mass
$m_{{\tilde \chi}^{0}_{i}}$  for  input parameters of scenario
$Sc_4$ with $\Phi_L=0.$  Here, according Eq. \eqref{eq:normeML}, the
graphs with the $+$ and $-$ signs  are represented in  solid and
dashed lines, respectively.} \label{fig:MLvsmj1}
\end{figure}

\begin{figure} \centering
\begin{picture}(31.5,21)
\put(1,2){\includegraphics[width=70mm]{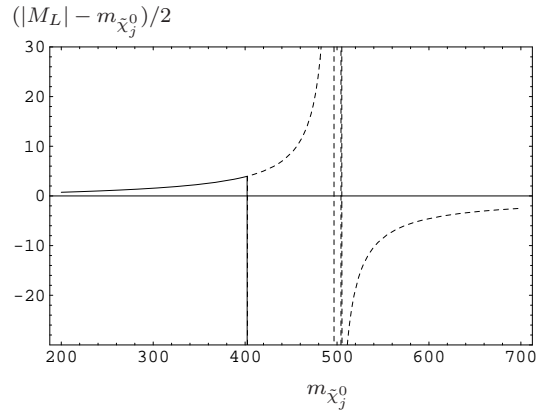}}
\put(15.75,0.7){\scriptsize{$m_{{\tilde \chi}^{0}_{j}}$}}
\put(1.05,19.25){\scriptsize{$(|M_L| - m_{{\tilde
\chi}^{0}_{j}})/2$}}
\end{picture}
\caption{The same as in Fig. \ref{fig:MLvsmj1} but with
$\Phi_L=\pi.$}
 \label{fig:MLvsmj2}
\end{figure}

\section{Determining L-R SUSY parameters}
\label{sec-detparameters} In this section we investigate the
behavior of  $|M_L |$ and $\Phi_L$   when the eigenphases $\eta_j,
\, j=1,2,$ change. We concentrate in two possible scenarios $Snc_2$
and $Snc_3,$ described in Table \ref{tab:tablados}, for fixed input
constants $g_L=g_R=g_V=0.65$ and $k_u=92.75.$ Thus, in  these cases
we assume that the physical masses  $ m_{{\tilde \chi}^{0}_{1}} $
and $ m_{{\tilde \chi}^{0}_{2}}$ as well as  $\mu$ and $M_R$ are
known. The $M_V$ parameter would eventually  be allowed to vary but
in this case we assume that  it has a fixed value in each one of the
mentioned scenarios.

 Figure \ref{fig:MLvsEtaja2} shows the behavior of the norm of
 $M_L,$ calculated from Eq. \eqref{genMLformula},
 as a function of the eigenphase $\eta_1,$  with input parameters
 of scenario $Snc_2$  when $m_{{\tilde \chi}^{0}_{1}}=164.36 \, {\rm GeV}.$
This is a scenario, similar to  the  Sp1-type scenario used in
\cite{key2} in the context of the MSSM,  characterized by a big rate
between $k_u$ and $k_d,$ i.e.,  $\tan\theta_k=30.$ For small values
of $\Phi_\mu,$ we observe small differences among the plots of
$|M_L|.$  For all the plots shown in this Figure, the mean value of
$|M_L|$ is $165.75$GeV approximately and the maximum amplitude
difference of them is $0.6$GeV approximatively.

Figures \ref{fig:MLvsEtajb2} and \ref{fig:MLvsEtajc2} show the
dependence of the mixing phase $\Phi_L$ and the relative phase
$\phi_L - {\rm Arg}(\eta_1),$ respectivement, calculated from Eq.
\eqref{genMLformula}, with respect to the eigenphase $\eta_1,$ in
the $Snc_2$ scenario with  $m_{{\tilde \chi}^{0}_{1}}=164.36 \, {\rm
GeV}.$ We observe, for all the cases $\Phi_\mu =0, {\pi \over
8},{\pi \over 6},\pi,$ a linear dependence between $\Phi_L$ and
${\rm Arg}(\eta_1).$ Thus,  $\Phi_L \approx {\rm Arg}(\eta_1) $ when
${\rm Arg}(\eta_1) \in \left[-{\pi\over2}, {\pi \over 2}\right],$
$\Phi_L \approx \pi + {\rm Arg}(\eta_1) $ when ${\rm Arg}(\eta_1)
\in \left(-\pi, - {\pi \over 2}\right)$ and $\Phi_L \approx - \pi +
{\rm Arg}(\eta_1) $  when ${\rm Arg}(\eta_1) \in \left( {\pi \over
2}, \pi \right).$

 Figure \ref{fig:MLvsEtaja1} shows the behavior of the norm of
 $M_L,$ calculated from Eq. \eqref{genMLformula},
 as a function of the eigenphase $\eta_2,$  with input parameters
 of scenario $Snc_2$  when $m_{{\tilde \chi}^{0}_{2}}=241.94 \, {\rm GeV}.$
 In this case, the  mean amplitude difference  of  $|M_L|$ for the different plots
 is greater than before, $120$GeV approximately. However,
 in the region of small  $\Phi_\mu,$  and  Arg$(\eta_j), j=1,2,$  the  results are
 closely similar, that is,  the values of  $|M_L|$ concentrate in the
 range $150{\rm GeV}-170{\rm GeV}.$  Figures
 \ref{fig:MLvsEtajb1} and \ref{fig:MLvsEtajc1}  show the
dependence of the mixing phase $\Phi_L$ and of the relative phase
$\phi_L - {\rm Arg}(\eta_2),$ respectivement, calculated from Eq.
\eqref{genMLformula}, with respect to the eigenphase $\eta_2,$ for
input parameters of  scenario $Snc_2$. As before,  for all the cases
$\Phi_\mu =0, {\pi \over 8},{\pi \over 6},\pi,$ we observe the same
linear dependence between $\Phi_L$ and ${\rm Arg}(\eta_1)$
practically. However, the non exact linearity implies  differences
in  the behavior  of $|\Phi_L|$ when it is  measured with respect to
Arg$(\eta_2)$ and $\Phi_L,$ as we can see by comparing Figures
\ref{fig:MLvsEtaja1} and \ref{fig:MLvsFLcompEtaj}.

\begin{table}
\begin{center}
\begin{tabular}{c c c c c c c c}\cline{1-8}\\
Scenario & $|\mu| \;$ & $ \quad \Phi_u \quad $ &$ m_{{\tilde
\chi}^{0}_{1}} \; $ & $m_{{\tilde \chi}^{0}_{2}} \; $ & $ M_R \;$ &
$ M_V \;$ & $ \tan\theta_k  $ \\
\\\cline{1-8} \\
$Snc_2$& 248 & \parbox[c]{0.5cm}{$ 0$  \\ ${\pi /  8}$ $ {\pi / 6 }$
$\pi $} & $ 164.36$ & $241.94$ & $300$ & $20$ & $30 $
\\ \\  \hline \\ $Snc_3 $& 150 & \parbox[c]{0.5cm}{$ 0$   ${\pi / 8}$ $ {\pi / 6 }$ $\pi $} & $
156.24$ & $236.79$ & $300$ & $50$ & $4.0$
\\ \\ \hline
\end{tabular}
\end{center}
\caption{Input parameters for scenarios $Snc_2$ and $Snc_3.$ All
mass quantities are in GeV and all angles are in radians.}
\label{tab:tablados}
\end{table}

\begin{figure} \centering
\begin{picture}(31.5,21)
\put(1,2){\includegraphics[width=70mm]{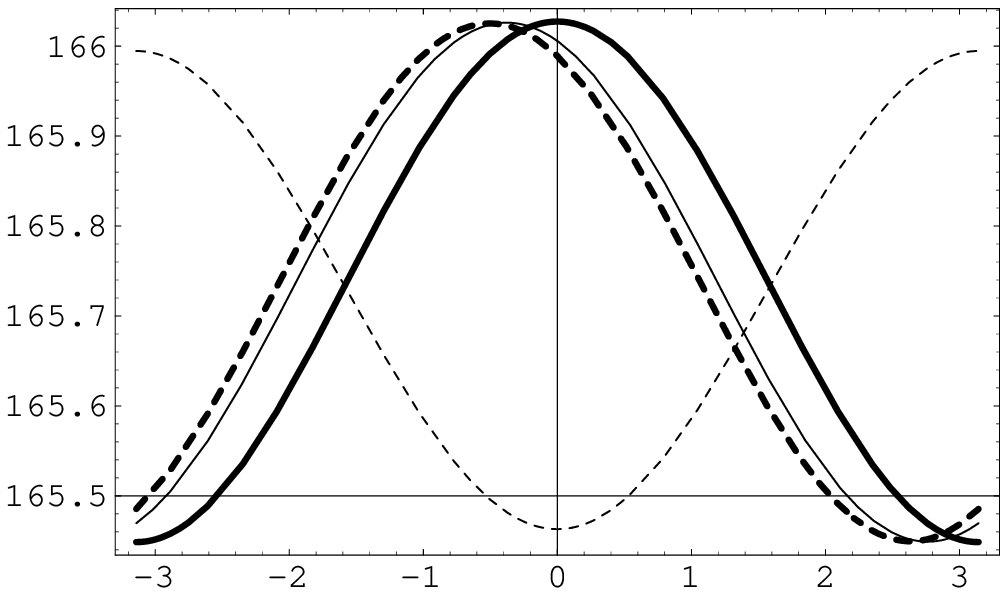}}
\put(15.75,0.7){\scriptsize{${\rm Arg}(\eta_1) \, {\rm (rad)}$}}
\put(1.05,19.25){\scriptsize{$|M_L| \, {\rm (GeV)} $}}
\end{picture}
\caption{Norm of $M_L$  as a  function of ${\rm Arg}(\eta_1),$
according to scenario $Snc_2,$  for $m_{{\tilde
\chi}^{0}_{1}}=164.36 \, {\rm GeV},$  $\Phi_{\mu}= 0 \; ({\rm heavy
\; solid}),$ $ \pi/8  \;  ({\rm light \; solid }),$$ \pi /6 \; ({\rm
heavy \; dashed}),$ $ \pi \; ({\rm light \; dashed}).$}
\label{fig:MLvsEtaja2}
\end{figure}

\begin{figure} \centering
\begin{picture}(31.5,21)
\put(1,2){\includegraphics[width=70mm]{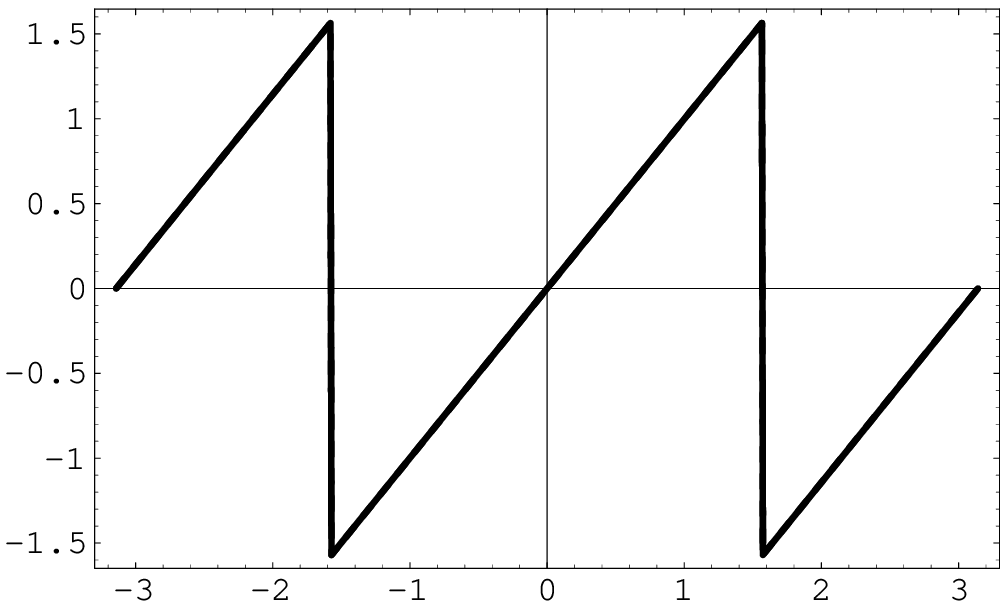}}
\put(15.75,0.7){\scriptsize{${\rm Arg} (\eta_1) \, {\rm (rad)}$}}
\put(1.05,19.25){\scriptsize{$ \Phi_L \, {\rm (rad)} $}}
\end{picture}
\caption{Mixing parameter $\Phi_L$ as a function of ${\rm Arg}
(\eta_1)$ with the same set of fixed parameters used in Fig.
\ref{fig:MLvsEtaja2}. } \label{fig:MLvsEtajb2}
\end{figure}

\begin{figure} \centering
\begin{picture}(31.5,21)
\put(1,2){\includegraphics[width=70mm]{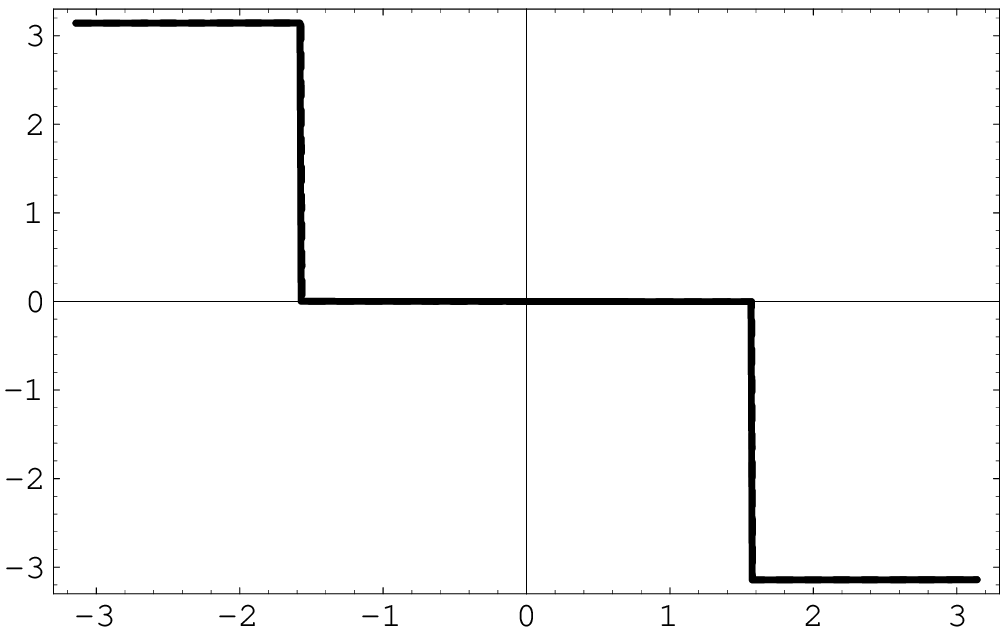}}
\put(15.75,0.7){\scriptsize{${\rm Arg}(\eta_1)\, {\rm (rad)}$}}
\put(1.05,19.25){\scriptsize{$[\Phi_L - {\rm Arg} (\eta_1)] \, {\rm
(rad)} $}}
\end{picture}
\caption{Relative difference between $\Phi_L$ and ${\rm Arg}
(\eta_1)$ as a function of ${\rm Arg} (\eta_2),$ as observed from
Fig. \ref{fig:MLvsEtajb2}.} \label{fig:MLvsEtajc2}
\end{figure}

\begin{figure} \centering
\begin{picture}(31.5,21)
\put(1,2){\includegraphics[width=70mm]{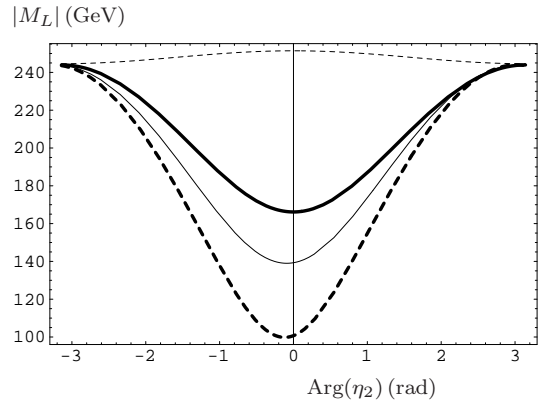}}
\put(15.75,0.7){\scriptsize{${\rm Arg}(\eta_2) \, {\rm (rad)}$}}
\put(1.05,19.25){\scriptsize{$|M_L| \, {\rm (GeV)} $}}
\end{picture}
\caption{Norm of  $|M_L|$  as  a function of $\eta_2$ in the case of
scenario $Snc_2,$ with  $m_{{\tilde \chi}^{0}_{2}} = 241.94 \, {\rm
GeV},$  $\Phi_{\mu}=0 \; ({\rm heavy \; solid}),$ $ \pi/8 \; ({\rm
light \; solid }), $ $ \pi /6 \; ({\rm heavy \; dashed}), $ $ \pi \;
({\rm light \; dashed}).$} \label{fig:MLvsEtaja1}
\end{figure}

\begin{figure} \centering
\begin{picture}(31.5,21)
\put(1,2){\includegraphics[width=70mm]{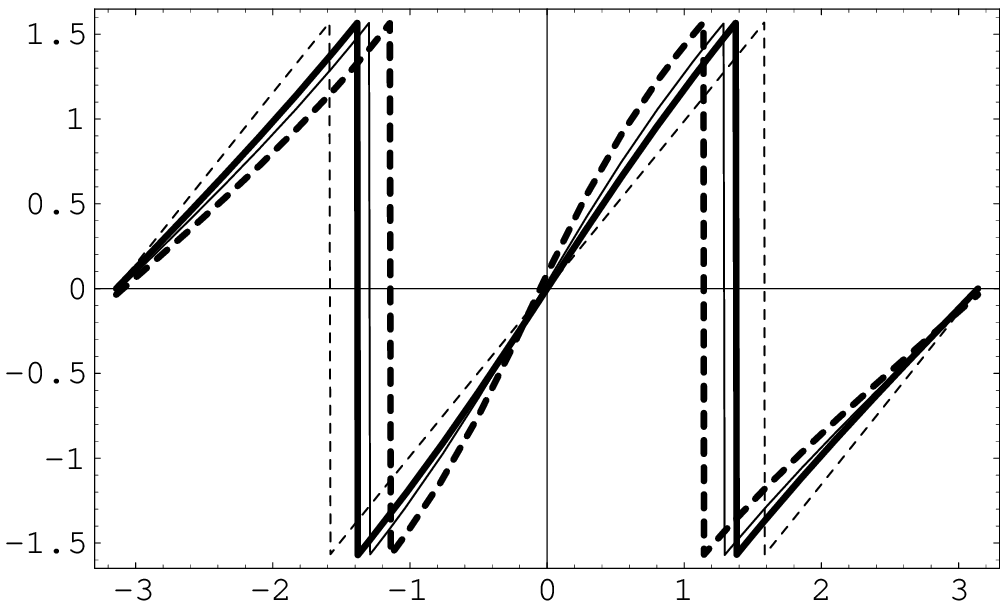}}
\put(15.75,0.7){\scriptsize{${\rm Arg} (\eta_2) \, {\rm (rad)}$}}
\put(1.05,19.25){\scriptsize{$ \Phi_L \, {\rm (rad)} $}}
\end{picture}
\caption{Mixing parameter $\Phi_L$ as a function of ${\rm Arg}
(\eta_2)$ with the same set of fixed parameters used in Fig.
\ref{fig:MLvsEtaja1}. } \label{fig:MLvsEtajb1}
\end{figure}

\begin{figure} \centering
\begin{picture}(31.5,21)
\put(1,2){\includegraphics[width=70mm]{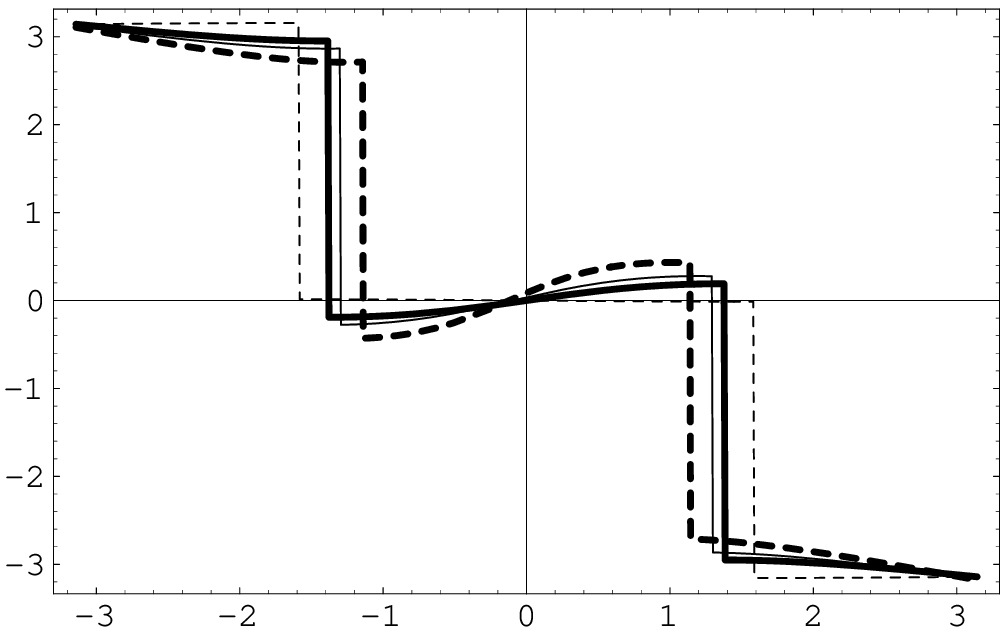}}
\put(15.75,0.7){\scriptsize{${\rm Arg}(\eta_2)\, {\rm (rad)}$}}
\put(1.05,19.25){\scriptsize{$[\Phi_L - {\rm Arg} (\eta_2)] \, {\rm
(rad)} $}}
\end{picture}
\caption{Relative difference between $\Phi_L$ and ${\rm Arg}
(\eta_2)$ as a function of ${\rm Arg} (\eta_2),$ as observed from
Fig. \ref{fig:MLvsEtajb1} .} \label{fig:MLvsEtajc1}
\end{figure}

\begin{figure} \centering
\begin{picture}(31.5,21)
\put(1,2){\includegraphics[width=70mm]{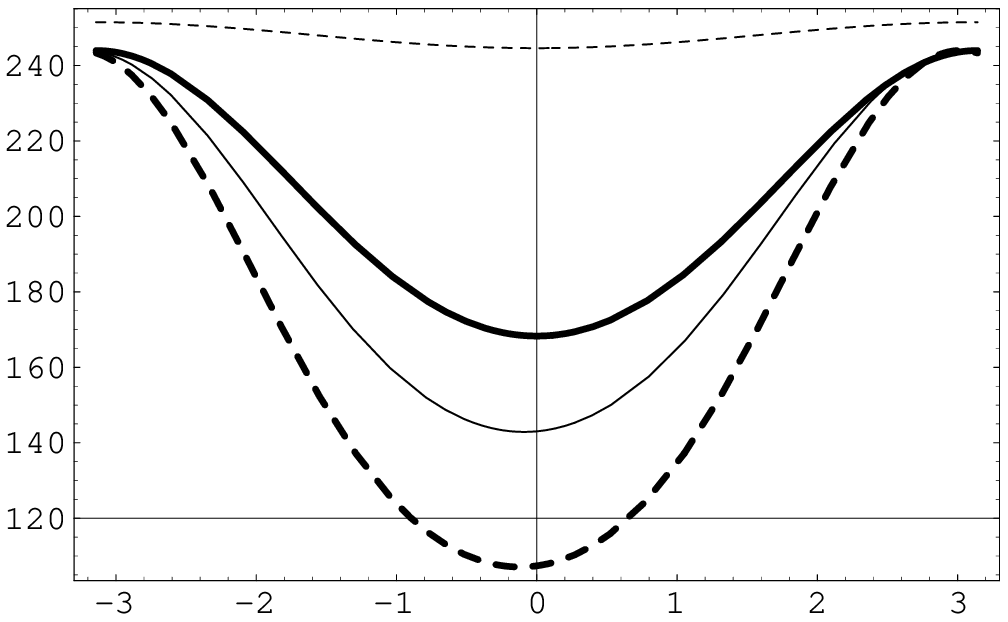}}
\put(15.75,0.7){\scriptsize{$\Phi_L \, {\rm (rad)}$}}
\put(1.05,19.25){\scriptsize{$|M_L| \, {\rm (GeV)} $}}
\end{picture}
\caption{Norm of  $|M_L|$  as  a function of $\Phi_L,$ computed from
Eq. \eqref{eq:normeML}, for  scenario $Snc_2,$ with $m_{{\tilde
\chi}^{0}_{2}} = 241.94 \, {\rm GeV},$   $\Phi_{\mu}=0 \; ({\rm
heavy \; solid}), $ $\pi/8 \;  ({\rm light \; solid }), $ $ \pi /6
\; ({\rm heavy \; dashed}), $ $\pi \; ({\rm light \; dashed}).$}
\label{fig:MLvsFLcompEtaj}
\end{figure}

\begin{figure}[t] \centering
\begin{picture}(31.5,21)
\put(1,2){\includegraphics[width=70mm]{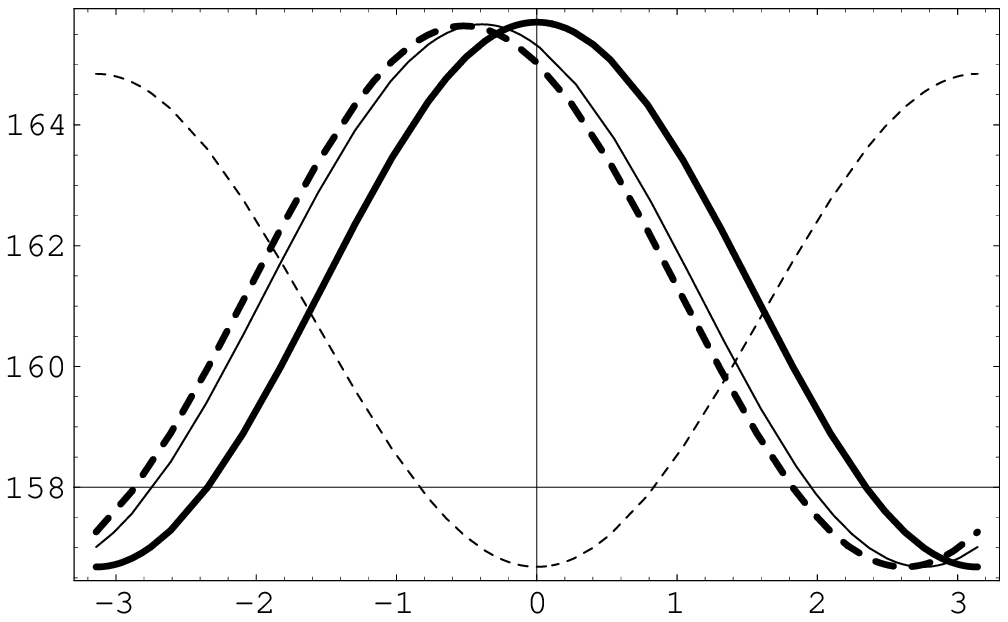}}
\put(15.75,0.7){\scriptsize{$\Phi_L \, {\rm (rad)}$}}
\put(1.05,19.25){\scriptsize{$|M_L| \, {\rm (GeV)} $}}
\end{picture}
\caption{Norm of  $|M_L|$  as  a function of $\Phi_L,$ computed from
Eq. \eqref{eq:normeML}, for  scenario $Snc_3,$ with $ m_{{\tilde
\chi}^{0}_{1}} = 156.24 \, {\rm GeV},$ $\Phi_{\mu}=0  \; ({\rm heavy
\; solid}),$ $ \pi/8 \; ({\rm light \; solid }), $ $ \pi /6 \; ({\rm
heavy \; dashed}), $ $ \pi \; ({\rm light \; dashed}).$}
\label{fig:MLvsFL1}
\end{figure}

\begin{figure} \centering
\begin{picture}(31.5,21)
\put(1,2){\includegraphics[width=70mm]{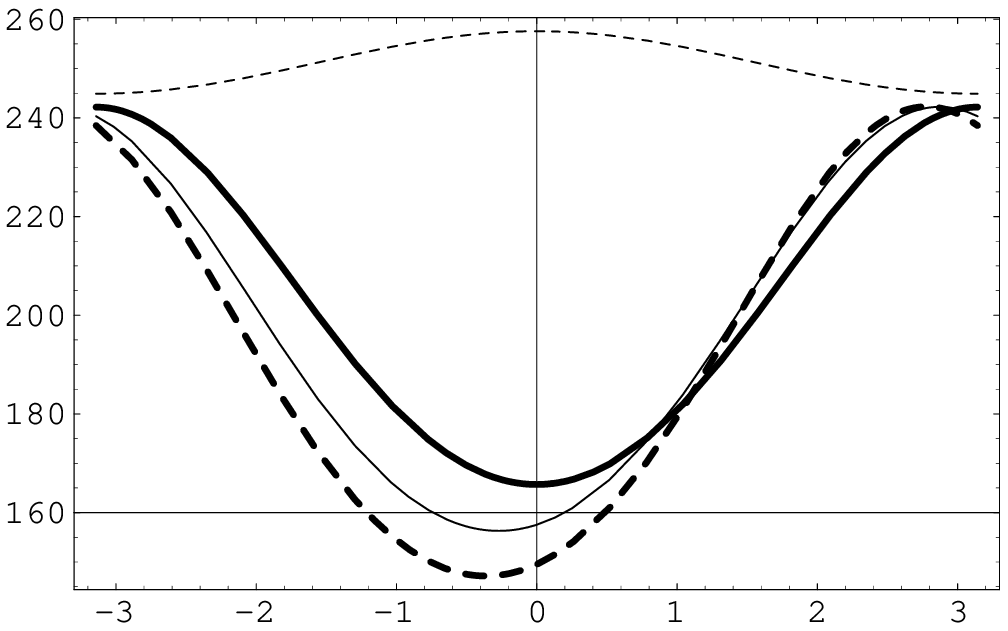}}
\put(15.75,0.7){\scriptsize{$\Phi_L \, {\rm (rad)}$}}
\put(1.05,19.25){\scriptsize{$|M_L| \, {\rm (GeV)} $}}
\end{picture}
\caption{The same inputs as in  Fig. \ref{fig:MLvsFL1}, but
considering the neutralino mass  $m_{{\tilde \chi}^{0}_{2}}=236.79
\,{\rm GeV}.$ } \label{fig:MLvsFL2}
\end{figure}

Let us now  assume  other possible scenario, $Snc_3,$ described in
Table \ref{tab:tablados},  where   $|\mu|=150$GeV  and
$\tan\theta_k=4.$ In this case, either the physical mass are given
by $m_{{\tilde \chi}^{0}_{1}} = 156.238$GeV or $ m_{{\tilde
\chi}^{0}_{2}} = 236.39$GeV, the same as in the case of scenario
$Snc_2,$ there exist practically a linear dependence between the
eigenphase and the mixing phase $\Phi_L.$ Thus, a description of
$|M_L|$ in terms of $\Phi_L$ is similar to the one based on the
eigenphases. Figure \ref{fig:MLvsFL1} shows the behavior of $M_L$
with respect to the phase $\Phi_L,$ computed from Eq.
\eqref{eq:normeML}, according to $Snc_3$ scenario with $m_{{\tilde
\chi}^{0}_{1}} = 156.238$GeV. Comparing with Fig.
\ref{fig:MLvsEtaja2}, constructed in similar conditions according to
scenario $Snc_2,$  in this case we observe a greater dispersion of
the values of $|M_L|$ when  $\Phi_\mu$ vary. For small phases $- 1
\le \Phi_L \le 1 $  and  $0 \le \Phi_\mu \le {\pi\over 8},$ the
values of $M_L$  lies in the range  $163{\rm GeV}-166{\rm GeV},$
approximately.  Figure \ref{fig:MLvsFL2} shows the behavior of $M_L$
with respect to the phase $\Phi_L,$ computed from Eq.
\eqref{eq:normeML} for input parameters of scenario $Snc_3$ with
$m_{{\tilde \chi}^{0}_{2}} = 236.39$GeV. Similarly, in this case,
the values of $|M_L|$ in  the mentioned lies in the range $155{\rm
GeV} -185{\rm GeV},$ approximately. Thus, the value of $|M_L|$ must
be localized in the intersection of these regions,  i.e.  it is
determined more accurately in the case of scenarios where the mass
$m_{{\tilde \chi}^{0}_{1}}$ is a known quantity.


\section{Conclusions}
\label{sec-conclusions}
In this paper we have studied the
implications of a complex symmetric neutralino mass matrix in the
context of left-right SUSY model. This matrix was described by seven
real parameters $|M_L|,$ $\Phi_L,$ $|\mu|,$ $\Phi_\mu,$ $M_R,$ $M_V$
and $\tan\theta_k.$ To find analytical expressions for the physical
masses $m_{{\tilde \chi}^{0}_{j}}, \, j=1,\ldots, 4, $ of the
neutralinos and some connecting relations among the parameters, at
the tree level, we have diagonalized this matrix by constructing the
corresponding diagonalizing unitary matrix. The masses, obtained by
solving the associated characteristic polynomial to this problem,
have been ordered by sizes and plotted as a function of the Higgsino
parameter $|\mu|,$  and also as a function of the mixing  phases
$\Phi_\mu$ and $\Phi_L.$ In the CP-conserving case, when all except
the $\mu$ parameter were fixed, according  to the possible scenarios
studied in Ref. \cite{key5}, we observe that there is not
intersection between the different curves representing the behavior
of the neutralino masses a function of $\mu.$  In the CP-violating
case, considering two possible scenarios, similar to the previous
ones but where $|\mu|$ was fixed and  $\Phi_\mu$ and $\Phi_L $ were
allowed to vary, we observe that there is not overlapping between
the surfaces representing the behavior of the neutralino masses
$m_{{\tilde \chi}^{0}_{1}}$ and $m_{{\tilde \chi}^{0}_{2}}.$

The inverse problem consisting to determine the mixing parameters
$|M_L|$ and $\Phi_L$ in terms of the rest of fundamental parameters
have been solved using the projector formalism without appeal to the
Jarlskog' projector formula. In this way, the so-called reduced
projectors have been expressed essentially in terms of the minors of
the determinant of the matrix formed from the product between the
original mass matrix and its adjoint. Thus, the $M_L$ parameter has
been disentangled and expressed in terms of the eigenphases by
solving a simple linear algebraic equation, in contrast to the
standard treatment where you need to solve a system of six linear
equations with six unknowns (see Appendix \ref{sec-ml-appendix}).
Moreover, combining the novel definition of the reduced projectors
with the Jarlskog' formula and then  solving a quadratic algebraic
equation we have obtained a new formula expressing the norm of $M_L$
in terms of the mixing parameter $\Phi_L$ and of the rest of the
fundamental parameters. This last formula provide a description for
the behavior of  $|M_L|$  in terms of $\Phi_L$  equivalent to the
one in terms of the eigenphases.

In the treatment of the inverse problem, in the CP-violating case,
we have considered two scenarios, the first one similar to the
$Sp1-$type considered in \cite{key2} in the context  of the  MSSM,
characterized by a big rate  between $k_u$ and $k_d$ and the second
one characterized by a relatively small rate  between $k_u$ and
$k_d,$ with similar conditions to those studied in \cite{key12} but
adapted to the CP-violating case. In both scenarios, we have
observed that the value of $|M_L|$ can be determined more accurately
if we know the the mass of the lighter neutralino.

A similar analysis can be carried out for the chargino sector. This
sector is more difficult to treat using the projector technique
because the corresponding chargino mass matrix is not symmetric and
requires two unitary matrices to diagonalize it. This analysis is
underway and will be reported in a separate communication.

\section*{Acknowledgments}
N. Alvarez  M.  thank  to the members of the organism {\it Baobab
Familial} of  Montreal for valuable support.  A.  De la Cruz would
like to thank Mariana Frank  for enlightening discussions. The
authors would also like thank the referees by valuable lecture and
suggestions about this article that contribute to enhance the
contents and presentation of it.

\appendix

\section{The standard method}
\label{sec-ml-appendix}

In this section we demonstrate the equivalence between the method
implemented in the above section and the one using the Jarlskog's
formula \eqref{Jarlskog-uno}, or well  Eq. \eqref{eq:ptildej}. The
method using the Jarlskog's formula to express $M_L$ in terms of the
eigenphases and of the rest of the fundamental parameters has been
used  in reference Ref.[2], in the case of the MSSM.

Equation  \eqref{eq:etaj}, for fixed $j,$ represent a system of four
complex algebraic equations serving to determine the six fundamental
L-R SUSY parameters and corresponding  eigenphase and  physical
neutralino mass in terms of the reduced projectors.  The explicit
form of this system of equations is obtained by inserting Eq.
\eqref{eq:Imatrix} into Eq.  \eqref{eq:etaj}, we give \beqa
\label{eq:mj1}
 \eta_j \, m_{{\tilde \chi}^{0}_{j}} &=& M_L - M (\sin\theta_k p_{j3}^\ast - \cos\theta_k
 p_{j4}^\ast) \\ \label{eq:mj2} &=& {M_{RV} p_{j2}^\ast + 2 \kappa M (\sin\theta_k p_{j3}^\ast - \cos\theta_k
 p_{j4}^\ast) \over p_{j2}} \quad \; \\ \label{eq:mj3}&=& {M \sin\theta_k (2 \kappa
 p_{j2}^\ast  -1) - 2 \mu p_{j4}^\ast \over p_{j3}} \\\label{eq:mj4} &=& {M \cos\theta_k (1- 2 \kappa
 p_{j2}^\ast) - 2 \mu p_{j3}^\ast \over p_{j4}}.
\eeqa The inverses of these equations determines the fundamental
LRSUSY parameters in terms of $p_{j\alpha},$ that is \beqa
\label{eq:MLpj} M_L &=& \eta_j \, m_{{\tilde \chi}^{0}_{j}} + M
(\sin\theta_k p_{j3}^\ast - \cos\theta_k
 p_{j4}^\ast), \\ \label{eq:MRV}M_{RV} &=&   {p_{j2}  \eta_j \,
 m_{{\tilde \chi}^{0}_{j}} - 2 \kappa M (\sin\theta_k p_{j3}^\ast - \cos\theta_k
 p_{j4}^\ast) \over p_{j2}^\ast }, \\ \mu &=&{ M \over 2} {(\sin\theta_k p_{j4} + \cos\theta_k p_{j3}) (1 - 2
\kappa p_{j2}^\ast) \over |p_{j3}|^2 - |p_{j4}|^2}, \eeqa where the
complex neutralino mass of ${\tilde \chi}^0_j$ is given by \be
\eta_j \, m_{{\tilde \chi}^{0}_{j}} = - M  {(\sin\theta_k
p_{j3}^\ast + \cos\theta_k p_{j4}^\ast) (1 - 2 \kappa p_{j2}^\ast)
\over |p_{j3}|^2 - |p_{j4}|^2}. \ee Also, as $M_{RV}$ is a real
quantity, from \eqref{eq:MRV} we  obtain \be \tan\theta_k  =  - {
{\rm Im}[p_{j4} (p_{j2}^\ast)^2] + 2 \kappa ( |p_{j4}|^2 -
|p_{j2}|^2 - |p_{j3}|^2 ) {\rm Im}[p_{j4} p_{j2}^\ast] \over {\rm
Im}[p_{j3} (p_{j2}^\ast)^2] + 2 \kappa ( |p_{j3}|^2 - |p_{j2}|^2 -
|p_{j4}|^2 ) {\rm Im}[p_{j3} p_{j2}^\ast]}. \ee

The complex  reduced projectors $p_{j2}, p_{j3}$ and $p_{j4}$ can be
computed from (\ref{eq:mj2}-\ref{eq:mj4}) without considering an
explicit dependence of $|M_L|$ and $\Phi_L.$ Solving  this system,
equivalent to  six linear equations with  six  real unknowns, we get

\beqa \label{eq:explpj2}p_{j2}&=& {1 \over 2 \kappa} + (m_{{\tilde
\chi}^{0}_{j}}^2 - 4 |\mu|^2) {\cal Z}^\ast_j, \\\label{eq:explpj3}
p_{j3}&=& 2 \kappa M (m_{{\tilde \chi}^{0}_{j}} \eta_j^\ast
\sin\theta_k {\cal Z}_j + 2 |\mu| e^{i \Phi_\mu } \cos\theta_k {\cal
Z}_j^\ast), \qquad  \\ \label{eq:explpj4} p_{j4}&=& - 2 \kappa M
(m_{{\tilde \chi}^{0}_{j}} \eta_j^\ast \cos\theta_k {\cal Z}_j + 2
|\mu| e^{i \Phi_\mu } \sin\theta_k {\cal Z}_j^\ast), \qquad \eeqa
where \be {\cal Z}_j = {\cal Z}_{j1} - m_{{\tilde \chi}^{0}_{j}}
\eta_j  {\cal Z}_{j2}, \ee with \beqa\nonumber {\cal Z}_{j1}& =& {1
\over 2 \kappa {\cal D}_j} \bigl\{ (m_{{\tilde \chi}^{0}_{j}}^2-
M_{RV^2}) (m_{{\tilde \chi}^{0}_{j}}^2 - 4 |\mu|^2) \\&-& 4 \kappa^2
M^2 [m_{{\tilde \chi}^{0}_{j}}^2 + 2 M_{RV} |\mu| e^{i \Phi_\mu }
\sin(2\theta_k)]\bigr\} \eeqa and \be {\cal Z}_{j2}= - { 2 \kappa
M^2 \over {\cal D}_j} [M_{RV} + 2 \mu e^{i \Phi_\mu }
\sin(2\theta_k)]. \ee Thus inserting \eqref{eq:explpj3} and
\eqref{eq:explpj4} into \eqref{eq:MLpj}, we obtain \beqa \nonumber
M_L &=& m_{{\tilde \chi}^{0}_{j}} \eta_j + 2 \kappa M^2 \\ \nonumber
&\times &( m_{{\tilde \chi}^{0}_{j}} \eta_j {\cal Z}_j^\ast + 2
|\mu| e^{-i \Phi_\mu} \sin(2\theta_k) {\cal Z}_j) \\
\label{eq:MLAjBj} &=& A_j \eta_j + B_j, \eeqa where $A_j$ and $B_j$
are given in Eqs. \eqref{eq:AJ} and \eqref{eq:BJ}, respectively.

\end{document}